\documentclass[twocolumn]{aastex63}

\hypersetup{linkcolor=red,citecolor=green,filecolor=cyan,urlcolor=magenta}

\submitjournal{ApJS}

\shorttitle{AO 0235+164 optical variability}
\shortauthors{Roy et al.}
\graphicspath{{./}{figures/}}
\usepackage{graphicx}	
\usepackage{amsmath}	
\usepackage{amssymb}	
\usepackage{ulem}
\usepackage{booktabs}
\usepackage{mathrsfs}
\usepackage{upgreek}
\usepackage{url}

\begin{document}

\title{Study of variability in long-term multiwavelength optical lightcurves of blazar AO 0235+164}

\correspondingauthor{Abhradeep Roy}
\email{abhradeep.1996@gmail.com, abhradeep.roy@tifr.res.in}

\author[0000-0002-4998-4560]{Abhradeep Roy}
\affiliation{Department of High Energy Physics, Tata Institute of Fundamental Research, Homi Bhabha Road, Mumbai-400005, India}

\author[0000-0002-9331-4388]{Alok C. Gupta}
\affiliation{Aryabhatta Research Institute of Observational Sciences (ARIES), Manora Peak, Nainital 263001, India}
\affiliation{Key Laboratory for Research in Galaxies and Cosmology, Shanghai Astronomical Observatory, Chinese Academy of Sciences, Shanghai 200030, China}

\author[0000-0001-5046-7504]{Varsha R. Chitnis}
\affiliation{Department of High Energy Physics, Tata Institute of Fundamental Research, Homi Bhabha Road, Mumbai-400005, India}

\author[0000-0002-3866-2726]{Sergio A. Cellone}
\affiliation{Complejo Astron\'omico El Leoncito (CASLEO, CONICET-UNLP-UNC-UNSJ), San Juan, Argentina}
\affiliation{Facultad de Ciencias Astron\'omicas y Geof\'\i sicas, Universidad Nacional de La Plata, La Plata, Buenos Aires, Argentina}

\author[0000-0003-1784-2784]{Claudia M. Raiteri}
\affiliation{INAF-Osservatorio Astrofisico di Torino, Via Osservatorio 20, I-10025 Pino Torinese, Italy}

\author[0000-0002-5260-1807]{Gustavo E. Romero}
\affiliation{Instituto Argentino de Radioastronom\'\i a (CCT-La Plata, CONICET; CICPBA; UNLP), Buenos Aires, Argentina}
\affiliation{Facultad de Ciencias Astron\'omicas y Geof\'\i sicas, Universidad Nacional de La Plata, La Plata, Buenos Aires, Argentina}

\author[0000-0002-1029-3746]{Paul J. Wiita}
\affiliation{Department of Physics, The College of New Jersey, 2000 Pennington Rd., Ewing, NJ 08628-0718, USA}

\author[0000-0003-0881-9275]{Anshu Chatterjee}
\affiliation{Department of High Energy Physics, Tata Institute of Fundamental Research, Homi Bhabha Road, Mumbai-400005, India}

\author[0000-0002-2565-5025]{Jorge A. Combi}
\affiliation{Facultad de Ciencias Astron\'omicas y Geof\'\i sicas, Universidad Nacional de La Plata, La Plata, Buenos Aires, Argentina}
\affiliation{Instituto Argentino de Radioastronom\'\i a (CCT-La Plata, CONICET; CICPBA; UNLP), Buenos Aires, Argentina}
\affiliation{Deptamento de Ingenier\'ia Mec\'anica y Minera, Universidad de Ja\'en, Campus Las Lagunillas s/n Ed. A3 Ja\'en, 23071, Spain}

\author[0000-0002-9137-7019]{Mai Liao}
\affiliation{CAS Key Laboratory for Researches in Galaxies and Cosmology, Department of Astronomy, University of Science and Technology of China, Hefei, Anhui 230026, China}
\affiliation{School of Astronomy and Space Science, University of Science and Technology of China, Hefei, Anhui 230026, China}

\author[0000-0002-7559-4339]{Arkadipta Sarkar}
\affiliation{Deutsches Elektronen-Synchrotron, Platanenallee 6, D-15738 Zeuthen, Germany}

\author[0000-0003-1743-6946]{Massimo Villata}
\affiliation{INAF-Osservatorio Astrofisico di Torino, Via Osservatorio 20, I-10025 Pino Torinese, Italy}



\begin{abstract}

We present a long-term and intraday variability study on optical multiwaveband ($U\!BV\!RI$) data from the blazar AO\,0235+164 collected by various telescopes for $\sim$44 years (1975--2019). The blazar was found to be significantly variable over the years in all wavebands with a variation of about six magnitudes between its low and active states. The variations in the different wavebands are highly correlated without any time-lag. We did not observe any significant trend in color variation with time, but we observed a bluer-when-brighter trend between the $B-I$ color index and the $R$-magnitude. Optical $BV\!R$-band spectral energy distributions always show a convex shape. Significant intraday variability was frequently seen in the quasi-simultaneous observations of AO\,0235+164 made on 22 nights in $R$ and $V$-bands by the CASLEO and CAHA telescopes during 1999--2019. We also estimated the central supermassive black-hole mass of $7.9\times10^7 M_{\odot}$ by analyzing the broad Mg II emission line in AO\,0235+164's spectrum. We briefly explore the probable physical scenarios responsible for the observed variability.

\end{abstract}



\section{Introduction}
\label{sec:intro}


\begin{figure*}[ht]
    \centering
    \includegraphics[trim={3.2cm 4cm 3.8cm 4cm},clip, width=\textwidth]{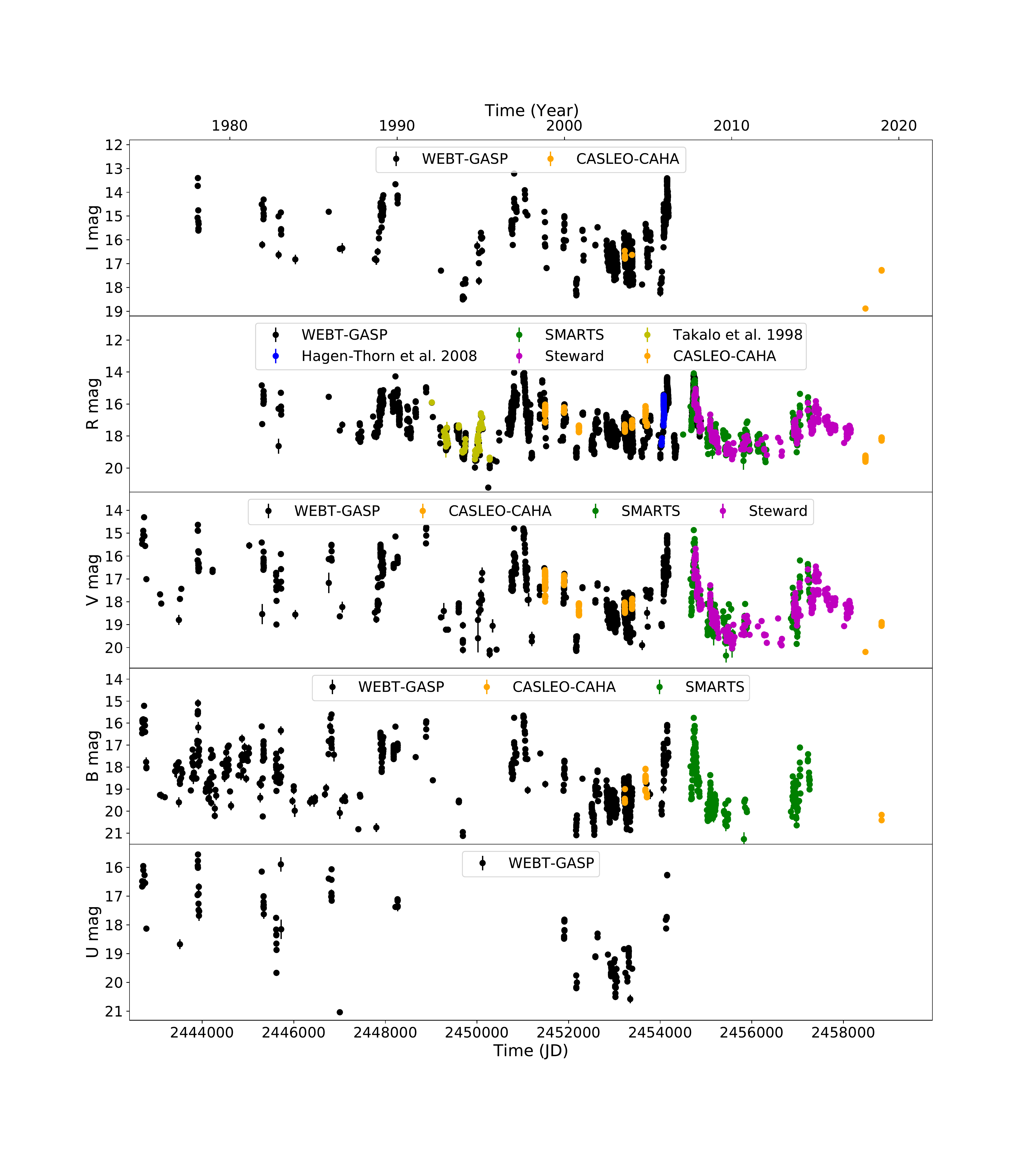}
    \caption{Long-term multiwavelength optical ($U$, $B$, $V$, $R$, $I$) lightcurves of AO\,0235+164 observed from multiple ground-based telescopes between JD 2442689 (1975 October 3) and JD 2458835 (2019 December 17).}
    \label{fig:mwlc}
\end{figure*}

\noindent
Blazars belong to the radio-loud (RL) class of active galactic nuclei (AGNs). This extremely variable class is the union of BL Lacertae objects (BL Lacs) and flat spectrum radio quasars (FSRQs). Blazars host a large-scale relativistic jet of plasma pointing very close to the observer's line of sight \citep{1995PASP..107..803U}. The jet is launched from the very near vicinity of the  supermassive black hole (SMBH) of mass 10$^{6}$ -- 10$^{10}$ M$_{\odot}$ at the center of the AGN \citep[e.g.,][]{2002ApJ...579..530W}. Blazars are characterized by highly variable emission throughout the whole electromagnetic (EM) spectrum, from radio to $\gamma$-rays, and their spectral energy distributions (SEDs) are characterized by two broad humps \citep{1998MNRAS.299..433F}. Blazars display high and variable polarization from radio to optical bands, and emit predominately non-thermal emission in the entire EM spectrum. The low-energy hump is ascribed to synchrotron radiation from relativistic leptons, whereas the high-energy hump arises from inverse Compton (IC) processes and sometimes from hadronic processes \citep[e.g.,][and references therein]{1983ApJ...264..296M,2003APh....18..593M,2017SSRv..207....5R}. \\
\\
Blazars display flux variability on diverse timescales ranging from a few minutes to several years. Blazar variability has often been divided into three categories, depending on the cadence of the observations: (i) microvariability \citep{1989Natur.337..627M},  or intraday variability  (IDV) \citep{1995ARA&A..33..163W}, or intra-night variability (INV) \citep{2004MNRAS.348..176S}, focusing on the variability over a day or less; (ii) short-term variability (STV), focusing on variability over days to weeks, (iii) and long-term variability (LTV), focusing on timescales of months to years \citep[e.g.][]{2004A&A...422..505G}. \\
\\
The BL Lac object AO\,0235+164
is at redshift $z =$ 0.94 \citep{1987ApJ...318..577C}. Optical spectroscopic and photometric observations of the object have discovered two foreground-absorbing systems at $z =$ 0.524 and $z =$ 0.851 \citep{1987ApJ...318..577C,1996A&A...314..754N,raiteri2007}. The flux of the source can be both absorbed and contaminated by these foreground systems, and the stars in them may act as gravitational micro-lenses that could contribute to the observed variability. \citet{1993ApJ...415..101A} did deep CFHT imaging of AO 0235+164 and reported that the source is weakly amplified by macrolensing / microlensing by stars in the foreground.\\
\\
AO\,0235+164 has been extensively observed in the past from radio to $\gamma$-ray bands either in individual EM bands or quasi-simultaneously in multiple EM bands and has shown variations in all those bands on diverse timescales \citep[e.g.,][and references therein]{1996ApJ...459..156M,1996A&A...310....1R,1998A&AS..129..577T,2000A&A...357...84Q,2000AJ....120...41W,2000A&A...360L..47R,2006A&A...459..731R,2008A&A...480..339R,2008ApJ...672...40H,2008AJ....135.1384G,2011ApJ...735L..10A,2012ApJ...751..159A,2017ApJ...837...45F,2018MNRAS.475.4994K,2020ApJ...902...41W}. It is one of the blazars which has displayed very high and variable optical/NIR polarization up to $\sim$45 percent \citep[e.g.,][and references therein]{1982MNRAS.198....1I,1993A&AS...98..393S,1999ApJS..121..131F, CRCM07, 2011PASJ...63..639I,2016ApJ...833...77I}.
In the Hamburg quasar monitoring program (HQM) this source was observed in the optical R band during 1988--1993, during which a 2.36$\pm$0.25 magnitude variation was detected; a particularly strong brightening in the source of $\sim$1.6 magnitude was reported during February 20--22, 1989 \citep{1994A&AS..106..349S}. In six nights of optical $B$ and $V$ bands observations during 21--27 September 1992, the blazar was found in an unusually bright state and IDV was detected in both $B$ and $V$ bands \citep{1996A&A...310....1R}. On another occasion, 6 nights of quasi-simultaneous $V$ and $R$ band observations in November 1999, revealed IDV with an amplitude of $\sim$100 percent over timescales of a day, while 0.5 magnitude changes were reported in both bands on a single night \citep{2000A&A...360L..47R}. In multicolor optical/NIR photometric ($BV\!RIJHK$) and $R$-band optical polarimetric observations of AO\, 0235+164  during its 2006 December outburst, variability on IDV timescales was detected, with increasing minimum timescale of variability from optical to NIR wavelengths; such variations were even detected in the optical polarization \citep{2008ApJ...672...40H}. In three nights of optical observations of the blazar in January -- March 2007, IDV and STV were detected \citep{2008AJ....135.1384G}. \\
\\
In quasi-simultaneous optical ($V$ and $R$ bands) and radio (22 GHz) observations of AO\,0235+164 during 1993--1996, the variability in optical bands showed amplitudes up to 1.5 magnitudes on STV timescales; although the radio variability is less dramatic, in general, it followed the optical behavior \citep{1998A&AS..129..577T}. For the 1997 AO\,0235+164 outburst, quasi-simultaneous multi-wavelength (MW) (radio, optical, NIR, and X-ray) observations were carried out. It was found that the source varied nearly simultaneously over 6 decades in frequency during the outburst and this result was explained in terms of a microlensing event \citep{2000AJ....120...41W}. \\
\\
An analysis of this source's variability over $\sim$25 years led to the suggestion of a $\sim$5.7 years quasi-periodicity of the main radio and optical flares \citep{2001A&A...377..396R}; however, the putative next outburst, predicted  to peak around February--March 2004, did not occur, and a new analysis of the optical light curves on a longer time span revealed a characteristic variability timescale of $\sim$8 years, which was also present in the radio data \citep{2006A&A...459..731R}. Recently, optical $R$ band photometric data taken during 1982--2019 showed 5 cycles of double-peaked periodicity of $\sim$8.13 years with a secondary peak following the primary one by $\sim$(1.5--2.0) years \citep{2022MNRAS.513.5238R}. In another MW campaign from radio to UV bands in 2006--2007, a huge NIR-optical-UV outburst with brightness increase of $\sim$5 magnitudes during February 19 -- 21, 2007 was detected \citep{2008A&A...480..339R}. During a  major outburst seen in 2009, changes in radio, optical, X-ray, and $\gamma$-ray bands were found to be strongly associated \citep{2011ApJ...735L..10A}. In another simultaneous MW observing campaign of this blazar between 2008 September and 2009 February, $\gamma$-ray activity was found to be well correlated with a series of NIR/optical flares, accompanied by an increase in the optical degree of polarization; the X-ray light curve showed a different 20-day high state of an unusually soft spectrum which did not match the extrapolation of the optical/UV synchrotron spectrum \citep{2012ApJ...751..159A}. \\
\\
AO 0235+164 is one of the sources that often used to be called OVV (optically violently variable). There are several such objects, like 3C 279, 3C 454.3, 4C 29.45, CTA 102, BL Lacertae, etc. Long-term achromaticity and zero lags have widely been found for these sources \citep{Bonning_2012,2021RAA....21..186Z,Fan_2006,2017Natur.552..374R,2015NewA...36....9G}. AO 0235+164 is peculiar because it is commonly considered a BL Lac, one of the furthest known, but it shares properties with FSRQs. It is also a complex source because its light is contaminated by the southern AGN, ELISA, and absorbed by an intervening galaxy. This paper has undertaken a detailed analysis of the source's optical brightness and spectral variability over a very long time span ($\sim$5 decades) as well as an investigation of its central engine. Our aim is to shed light on the long and short-term behavior of an emblematic BL Lac object through a detailed analysis of what is likely the most massive data set ever assembled for an object of this kind.
The paper is organized as follows. In \autoref{sec:obs}, we provide descriptions of the observations of AO\,0235+164. The \autoref{sec:ana} gives our data analysis methods and results. We present a discussion and conclusions in \autoref{sec:disc} and \autoref{sec:conc}, respectively.

\section{Observations} \label{sec:obs}

\noindent
Most of the optical $UBVRI$ observations of AO\, 0235+164 we have employed in this work are taken from The Whole Earth Blazar Telescope\footnote{\url{https://www.oato.inaf.it/blazars/webt}} (WEBT) \citep{2002A&A...390..407V,2017Natur.552..374R} which is an international collaboration of optical, near-infrared, and radio observers. WEBT has organized several monitoring campaigns on the blazar AO\, 0235+164, with the participation of many tens of observers and telescopes all around the world. Later, this source was studied by the WEBT and by its GLAST-AGILE Support Program (GASP) \citep{2008A&A...481L..79V,2009A&A...504L...9V}, which was started in 2007 to record quasi-simultaneous data of various blazars observed by the {\it AGILE} and {\it Fermi} (formerly {\it GLAST}) satellites. WEBT/GASP data on AO\, 0235+164 were published in \citet{2001A&A...377..396R,2005A&A...438...39R,2006A&A...459..731R,2008A&A...480..339R} and \citet{2012ApJ...751..159A}. \citet{2005A&A...438...39R} prescribed ways to remove the contribution of the southern galaxy ELISA from the observed optical flux densities and estimated the amount of absorption towards the source in excess of that from our Galaxy in  X-ray, ultraviolet, optical, and near-infrared bands.\\
\\
The WEBT and GASP data were calibrated following a common prescription, i.e., with the same photometry for the same reference stars. For calibration of the AO\, 0235+164 observations, the adopted photometric sequence includes stars 1, 2, and 3 from \citet{SBH85}. To build a reliable lightcurve for further analysis, clear outliers were removed and minor systematic offsets between various datasets were corrected.\\
\\
\noindent
AO\, 0235+164 was also observed with the 2.2\,m telescope of Calar Alto Astronomical Observatory (CAHA, Spain) in November -- December 2005, using the CAFOS instrument in imaging polarimetry mode, and photometric data were obtained by adding up the ordinary and extraordinary fluxes from each individual image \citep{CRCM07}. Photometric data were also obtained with the 2.15\,m telescope at Complejo Astron{\'o}mico El Leoncito (CASLEO, Argentina) along several runs in November 1999, December 2000, August 2004, and January 2005. Results from these data were published in \citet{1999A&AS..135..477R, 2000A&A...360L..47R, 2002A&A...390..431R} and in two papers by the WEBT collaboration focused on this blazar \citep{2005A&A...438...39R, 2006A&A...459..731R}. Data from a more recent (December 2019) observing run with the same telescope were used in \citet{2022MNRAS.513.5238R}. Magnitude calibration to the standard system was done using our own photometry of Landolt's (\citeyear{L09}) fields as well as standard stars in the field of AO\, 0235+164 \citep{SBH85, GKM01}.\\
\\
We also collected the publicly available optical $R$ and $V$-band data of AO\,0235+164, taken at Steward Observatory\footnote{\url{http://james.as.arizona.edu/~psmith/Fermi/DATA/Rphotdata.html}}, University of Arizona. These measurements employed  the 2.3 m Bok and 1.54 m Kuiper telescopes between 4 October 2008 and 12 February 2018, using the SPOL CCD Imaging/Spectropolarimeter attached to those two telescopes. Details about the instrument, observation, and data analysis are given in \citet{2009arXiv0912.3621S}. In addition, we included the optical-$BVR$ data from the Small and Moderate Aperture Research Telescope System (SMARTS) public archive\footnote{\url{http://www.astro.yale.edu/smarts/glast/home.php\#}}. The SMARTS consortium is part of the Cerro Tololo Inter-American Observatory (CTIO), Chile, and has been observing Fermi-Large Area Telescope (LAT)-monitored blazars in the optical $B$, $V$, $R$ and NIR $J$ and $K$ bands. Details about the SMARTS instruments, observations, and data analysis procedures are given in \citet{Bonning_2012}. These standard magnitudes observed by CASLEO, CAHA, SMARTS, and the Steward observatory were further corrected for the southern galaxy ELISA following \citet{2005A&A...438...39R}. We also added other $R$-band optical photometric data from the literature \citep{1998A&AS..129..577T,2008ApJ...672...40H}.

\section{Data analysis methods and results} \label{sec:ana}

\begin{figure*}
    \centering
    \includegraphics[width=\textwidth]{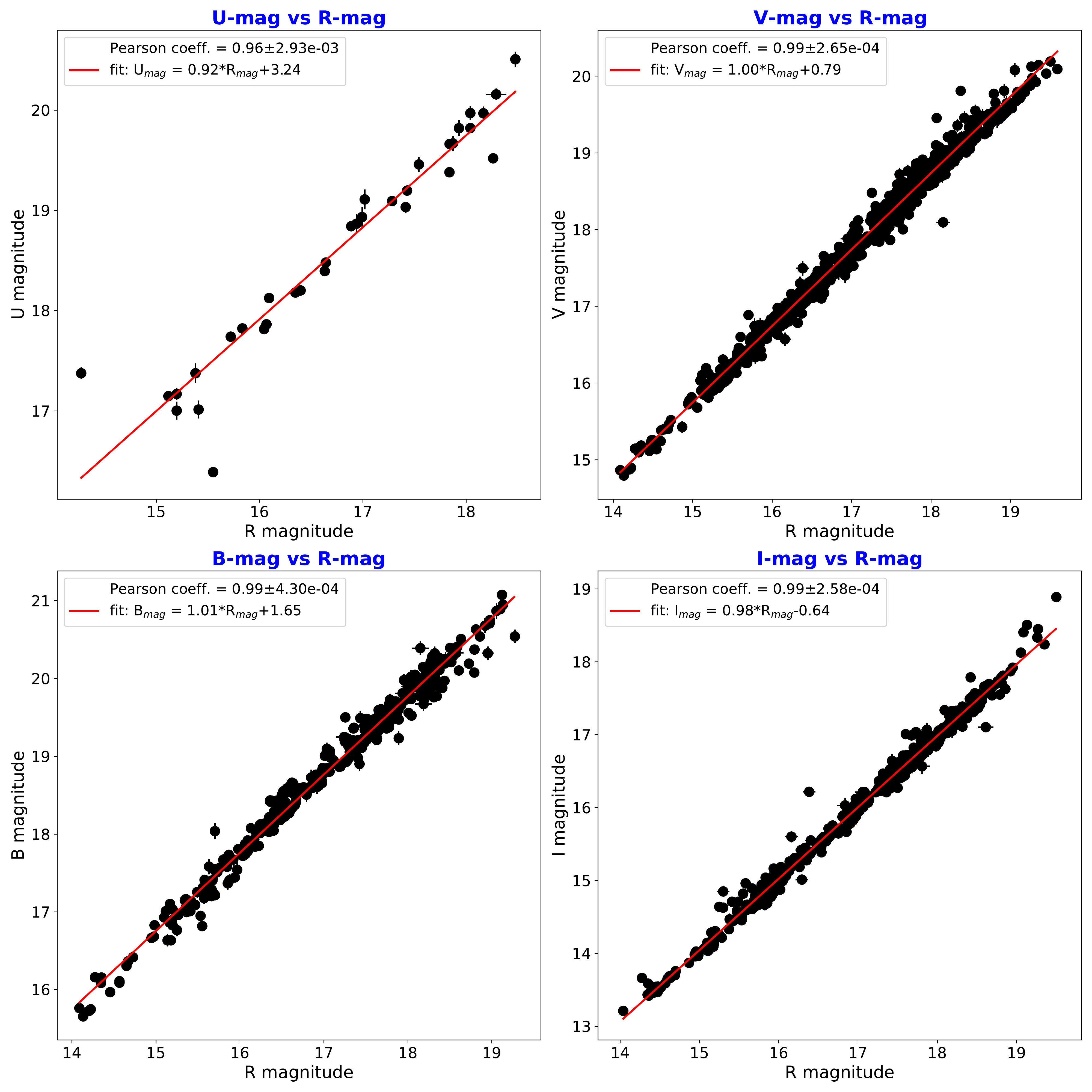}
    \caption{15-minute averaged $UBVI$ magnitudes versus $R$-magnitude plots for correlation study. $U$, $B$, $V$, and $I$-band observations show high linear correlation with $R$-band data. All the plots are fitted with straight lines.}
    \label{fig:corr}
\end{figure*}

\begin{figure*}
    \centering
    \includegraphics[width=\textwidth]{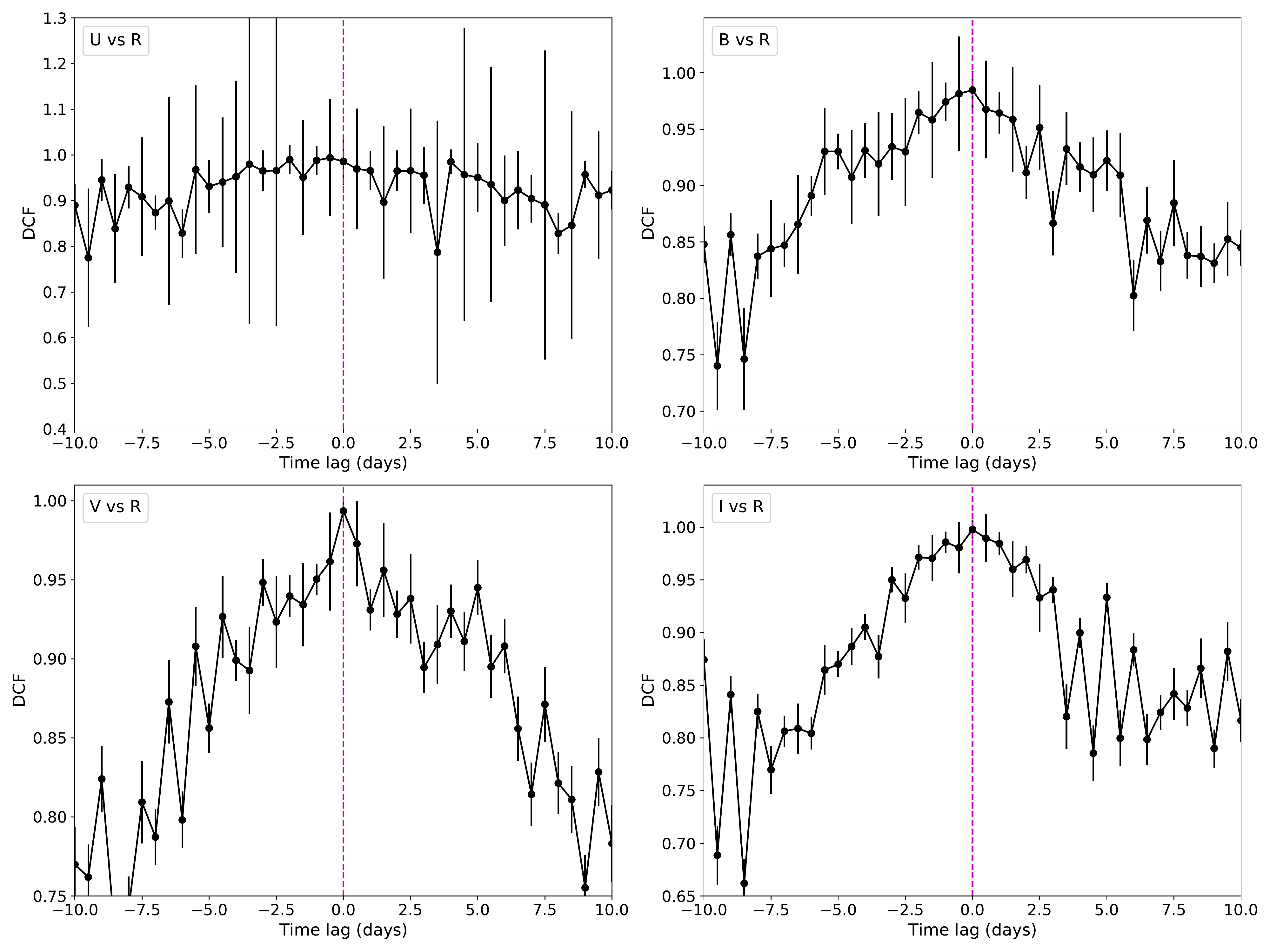}
    \caption{Results of discrete cross-correlation analysis of $U$, $B$, $V$, and $I$-band with respect to $R$-band in the full time range.}
    \label{fig:dcf}
\end{figure*}

\begin{figure*}[ht]
    \centering
    \gridline{\fig{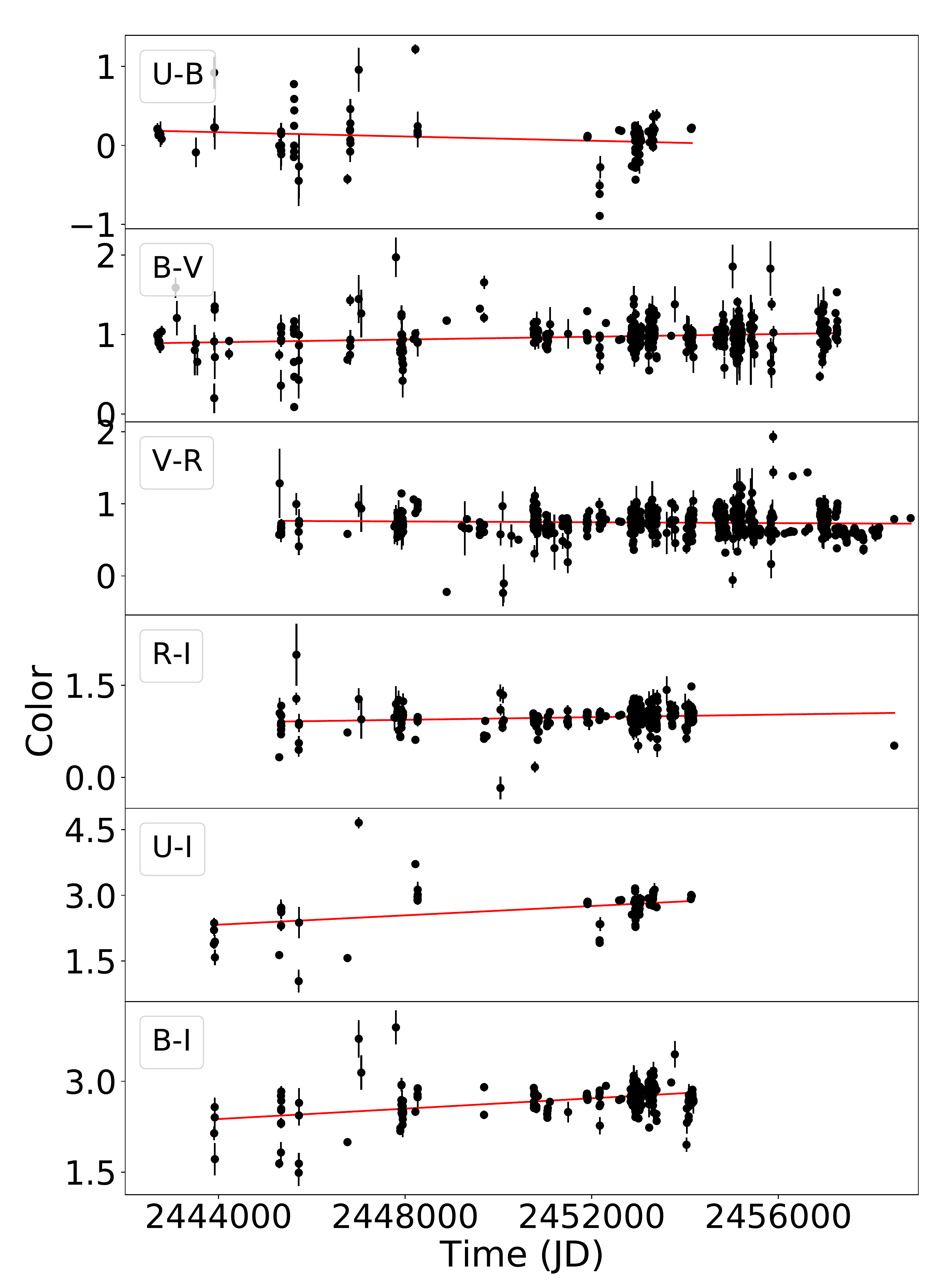}{0.5\textwidth}{(a)}
    \fig{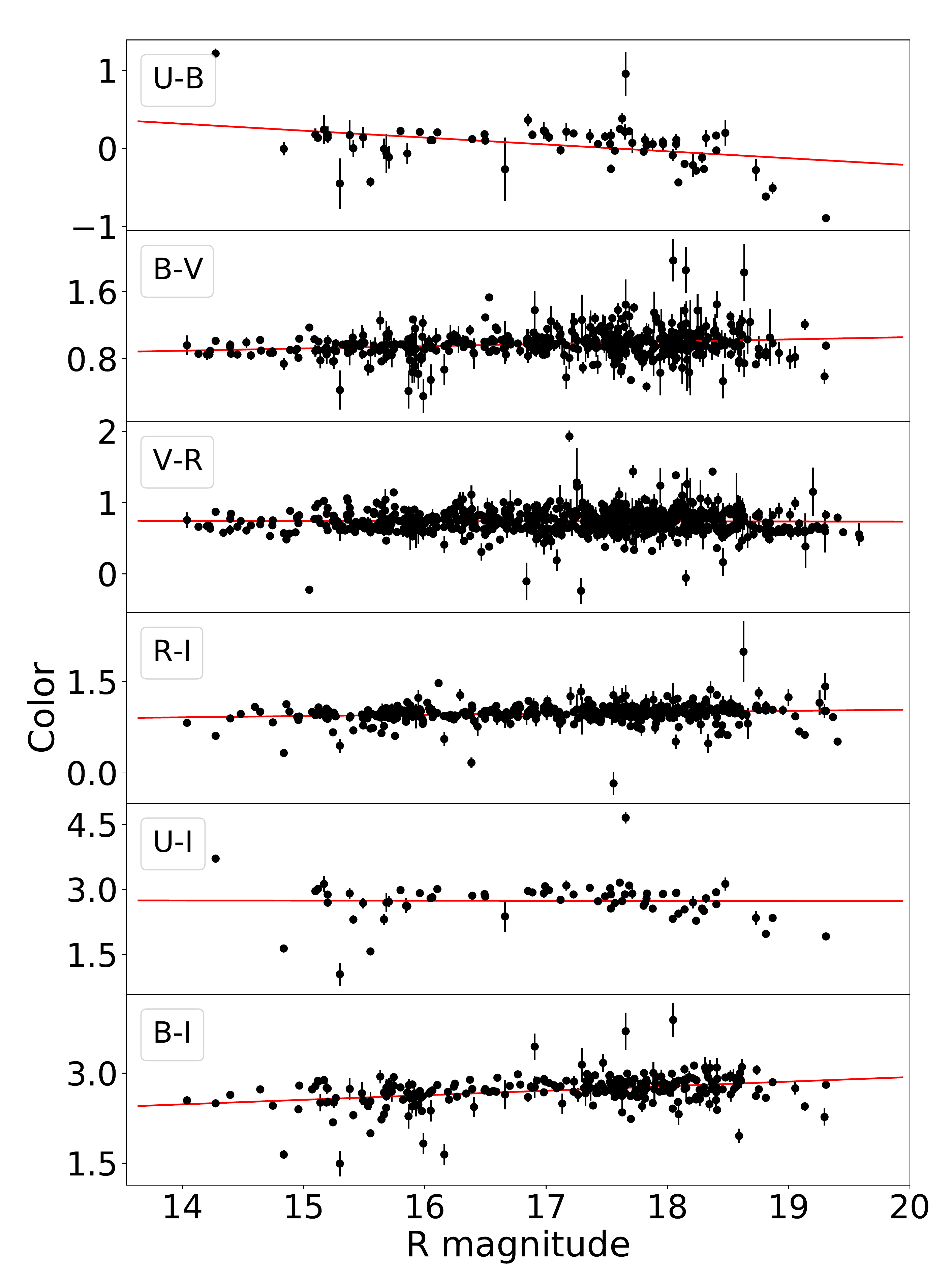}{0.505\textwidth}{(b)}
    }
    \caption{(a) Color variation with time. (b) Color variation with optical $R$ magnitude. The red line in each panel represents the straight line fit. Fit parameters are given in \autoref{tab:color_t} and \autoref{tab:color_r} respectively.}
    \label{fig:color_var}
\end{figure*}

\noindent
We combined all the optical $U$, $B$, $V$, $R$, $I$ band data to plot the long term (1974--2020) MW lightcurves of blazar AO\,0235+164 (\autoref{fig:mwlc}). We removed the observations with errors of more than 0.1 magnitudes and studied long-term and intraday variability, color variation, spectral properties, and inter-band correlations.

\subsection{Flux variability studies} \label{sec:ltv}

\noindent
We use different tools on the observed optical magnitudes to quantify the variability timescales and the corresponding significance in multiple optical wavebands. 

\begin{deluxetable}{cccccc}
\tablenum{1}
\tablecaption{Result of flux variability on optical $U\!BV\!RI$ long-term lightcurves of AO\,0235+164 \label{tab:ltvres}}
\tablewidth{0pt}
\tablehead{
\colhead{Optical} & \colhead{Total} & \colhead{$\chi^2_\mathrm{red.}$} &  \colhead{$\chi^2_{0.999, \mathrm{red.}}$} & \colhead{Status} & \colhead{Variability} \\ 
\colhead{filter} & \colhead{Obs.} & \colhead{} & \colhead{}  & \colhead{} & \colhead{amplitude (\%)} 
}
\startdata
$U$ & 109 & 904.5 & 1.47 & V & 548.8  \\
$B$ & 894 & 3246.7  & 1.15 & V & 590.9  \\
$V$ & 1403 & 5968.4  & 1.12 & V & 589.0  \\
$R$ & 5675 & 8715.5  & 1.06 & V & 718.8  \\
$I$ & 1173 & 3555.2  & 1.13 & V & 567.5  \\
\enddata
\tablecomments{In the fourth column 'V/NV' represents variable/non-variable status.}
\end{deluxetable}

\subsubsection{The \texorpdfstring{$\chi^2$}-test} \label{sec:chi2}

\noindent
For a time series of flux density observations, the $\chi^2$ is defined as,

\begin{equation}
    \chi^2 = \sum_{i=1}^N \frac{(\mathcal{M}_i-\mathcal{\Bar{M}})^2}{\varepsilon^2_i}
\end{equation}
where $\mathcal{M}_i$ is the magnitude obtained at the $i^{\text{th}}$ observation, $\varepsilon_i$ is the corresponding error in measurement, and $\mathcal{\Bar{M}}$ is the average magnitude. If the obtained $\chi^2$ value is higher than the critical $\chi^2$ value at 99.9 per cent significance level, we consider the source as variable. The critical value ($\chi^2_{0.999, d}$) depends on the degrees of freedom ($d$) of the dataset. The reduced $\chi^2$ values listed in \autoref{tab:ltvres} indicate that the source exhibits significant flux variations in all the optical wavebands.

\subsubsection{Variability amplitude} \label{sec:varamp}

\noindent
According to the relation given by \citet{1996A&A...305..42H}, we estimated the variability amplitudes ($V_M$) in percentage for the lightcurves in different wavelengths using the following formula,

\begin{equation}
\label{eq:vamp}
    V_M = 100 \times \sqrt{(\mathcal{M}_\mathrm{max}-\mathcal{M}_\mathrm{min})^2 - 2\,\Bar{\varepsilon}^2}\; (\%)
\end{equation}
where $\mathcal{M}_\mathrm{max}$ and $\mathcal{M}_\mathrm{min}$ are the maximum and minimum observed magnitude in a lightcurve, respectively, while $\Bar{\varepsilon}$ is the average error in magnitude measurements. We list the calculated variability of amplitudes in \autoref{tab:ltvres}.

\subsubsection{Correlation study} \label{sec:corr}

\noindent
To study the inter-band correlations, we first generated 15-minute binned optical $U\!BV\!RI$ lightcurves, and plotted the average $U$, $B$, $V$, and $I$-magnitudes against the average $R$-magnitudes for the time bins when the source was observed at both the wavebands (\autoref{fig:corr}). The magnitude-vs-magnitude plots show very good linear correlations. To take the uncertainty of magnitude measurements into account, we simulated 10000 datasets assuming that each magnitude measurement is Gaussian distributed. Then we calculated the mean and standard deviation of the Pearson correlation coefficients of all simulated datasets. We obtained high correlations ($>0.9$) with small uncertainties ($<0.003$) between all wavebands. \\
\\
Moreover, to find any time lag between the correlated optical lightcurves we computed the discrete correlation function (DCF) from the unbinned multiwavelength light curves, as the light curves consist of discrete data points. Following the method of \citet{1988ApJ...333..646E}, we computed the unbinned DCF (UDCF) between the $i^{\rm th}$ data point in one waveband ($a$) and the $j^{\rm th}$ data point in another ($b$) as

\begin{equation}
\label{eq:dcfeq}
    {\rm UDCF}_{ij} = \frac{(a_i-\bar{a})(b_j-\bar{b})}{\sigma_a \sigma_b},
\end{equation}
where $\bar{a}$ and $\bar{b}$ are the mean of the observed magnitudes, and $\sigma_a$ and $\sigma_b$ are the standard deviations of the corresponding datasets. Next, we calculated the discrete correlation function (DCF) at a certain time lag $\tau$ by averaging the ${\rm UDCF}_{ij}$s whose corresponding time lags $\Delta t_{ij} = t^a_i-t^b_j$ lie within the range [$\tau - \frac{\Delta \tau}{2}$, $\tau + \frac{\Delta \tau}{2}$] ($\Delta \tau$ is the time lag bin width), such that,

\begin{equation}
    {\rm DCF}(\tau) = \frac{1}{n}\sum {\rm UDCF}_{ij}(\tau).
\end{equation}
Following the suggestion of \citet{1994PASP..106..879W}, we computed the mean magnitudes ($\bar{a}$ and $\bar{b}$) and the standard deviations ($\sigma_a$ and $\sigma_b$) in \autoref{eq:dcfeq} using only those data points who fall within a given time lag bin, as the mean and standard deviation keep on changing for a time series originated from a stochastic process such as blazar emission. The error in the DCF($\tau$) computation in each bin is calculated as

\begin{equation}
    \sigma_{\rm DCF}(\tau) = \frac{1}{M-1}\sqrt{\sum_{k=1}^M ({\rm UDCF}_k - {\rm DCF}(\tau))^2}.
\end{equation}

\noindent
\autoref{fig:dcf} shows the DCFs of $UBVI$ bands with respect to the $R$-band observations. In all cases, the DCFs peak at zero time lag, except the $U$-band vs $R$-band DCF due to poor data sampling in the $U$-band. This explains the strong linearity in \autoref{fig:corr} and implies that the emission at all optical wavebands are coming from the same region in the jet and are produced from the same radiation mechanism.

\subsubsection{Color Variations} \label{sec:color}


\begin{figure*}
\begin{interactive}{animation}{AO0235p164_optical_SED_movie.mp4}
\includegraphics[width=\linewidth]{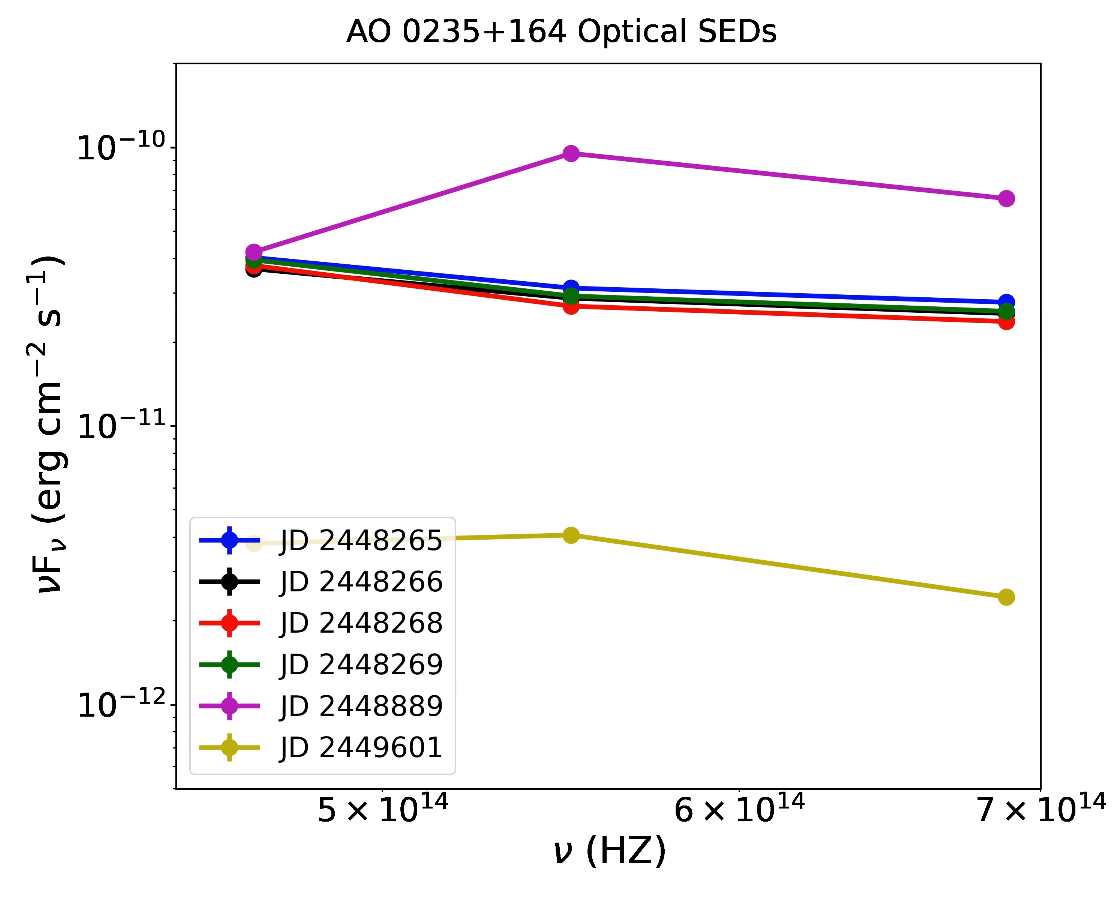}
\end{interactive}
\caption{An example frame of the AO 0235+164 optical SED animation that is available in the HTML version of this article. The duration of the animation is 1 minute and it contains a total of 360 one-day averaged optical SEDs, having 6 SEDs per frame. The observation dates of the SEDs are given in the plot legend.}
\label{fig:SEDmovie}
\end{figure*}

\begin{figure}
    \centering
    \includegraphics[width=\linewidth]{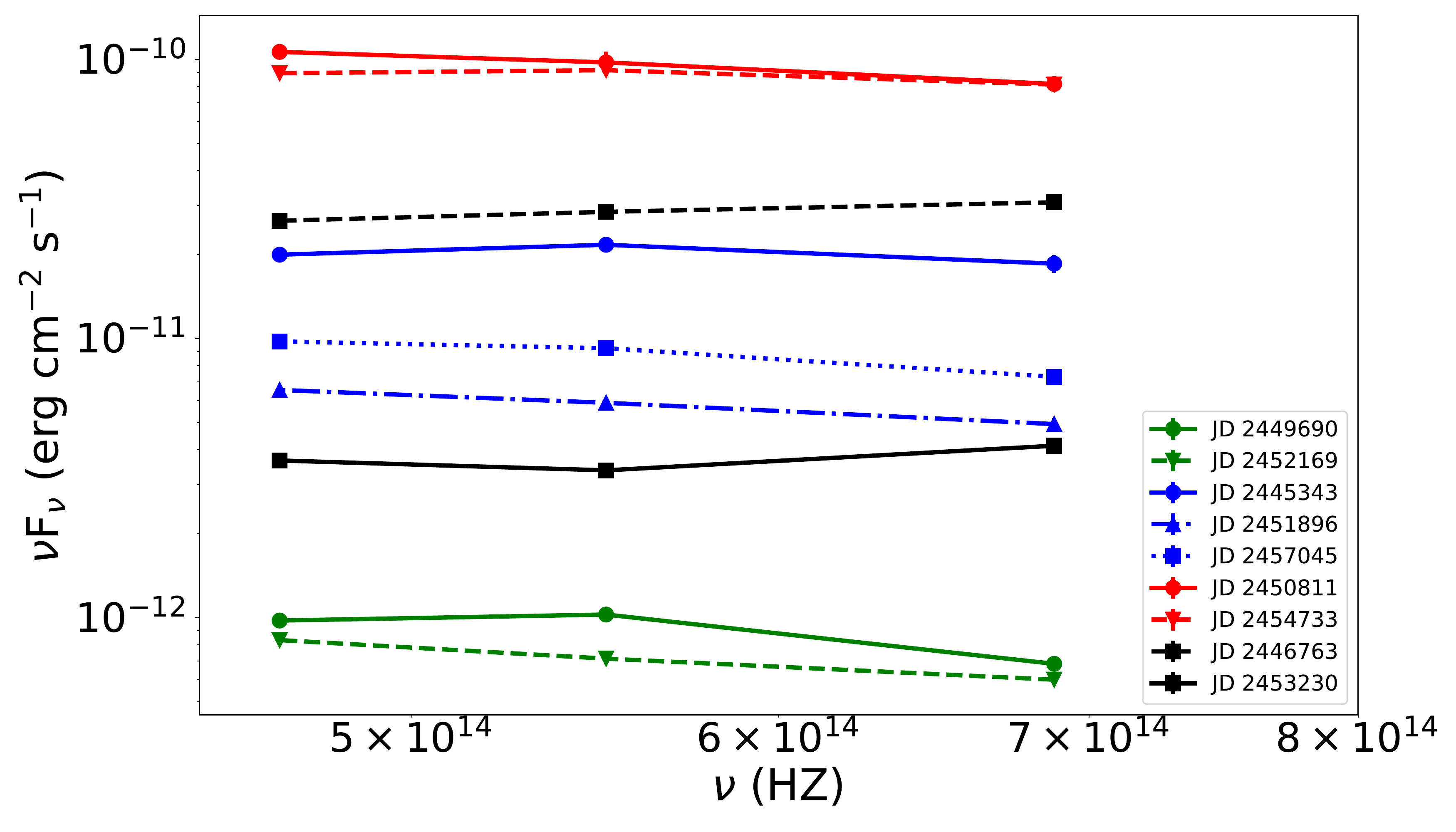}
    \caption{Examples of AO\,0235+164 optical intraday SEDs during three different states of brightness: (i) the green lines represent SED during quiescent states ($\nu$F$_{\nu}$ (erg cm$^{-2}$ s$^{-1}$) $<$ 10$^{-12}$), (ii) the blue lines show SED during moderately bright states (10$^{-12}$ $<$ $\nu$F$_{\nu}$ (erg cm$^{-2}$ s$^{-1}$) $<$ 3$\times$10$^{-11}$), (iii) the red lines show SED during outbursts ($\nu$F$_{\nu}$ (erg cm$^{-2}$ s$^{-1}$) $>$ 5$\times$10$^{-11}$). The black lines are examples of SED with spectral hardening on JD 2446763 and JD 2453230.}
    \label{fig:spec}
\end{figure}

\begin{figure*}[ht]
    \gridline{\fig{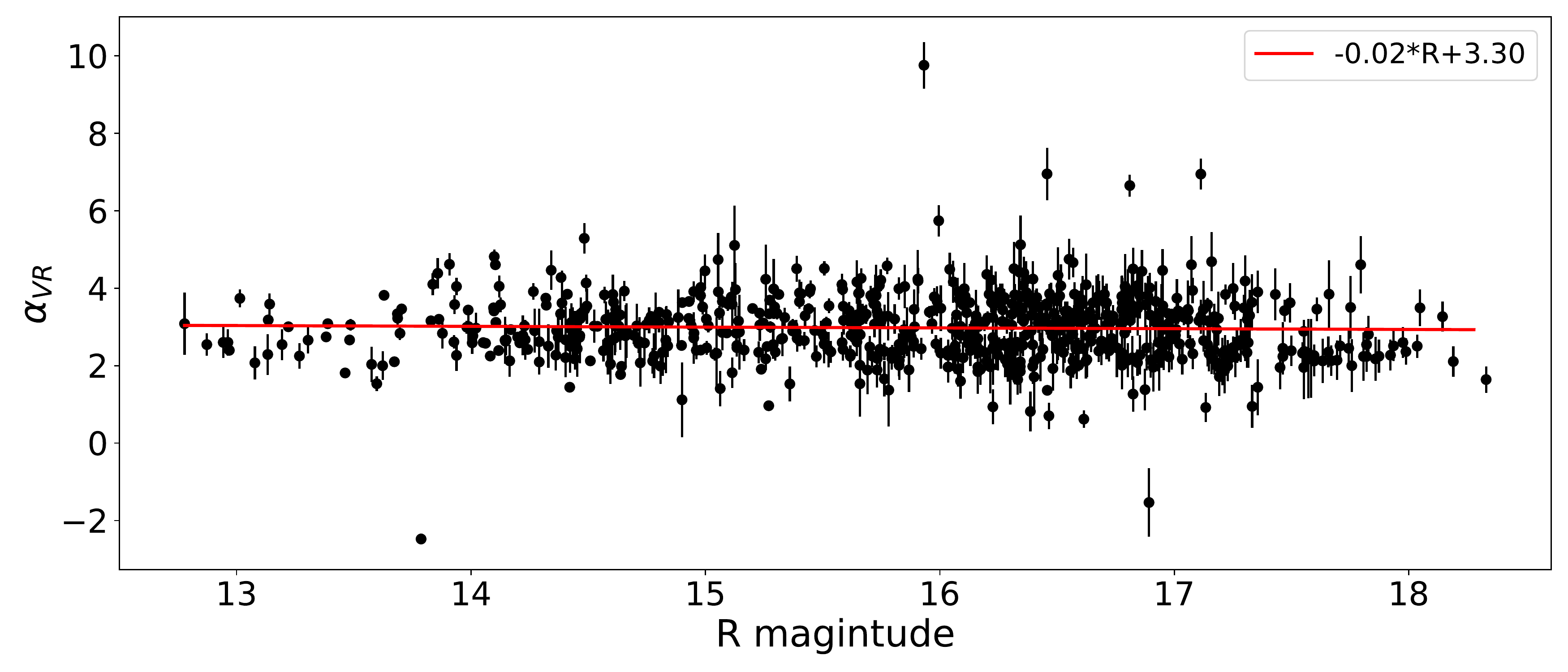}{0.5\textwidth}{(a)}
    \fig{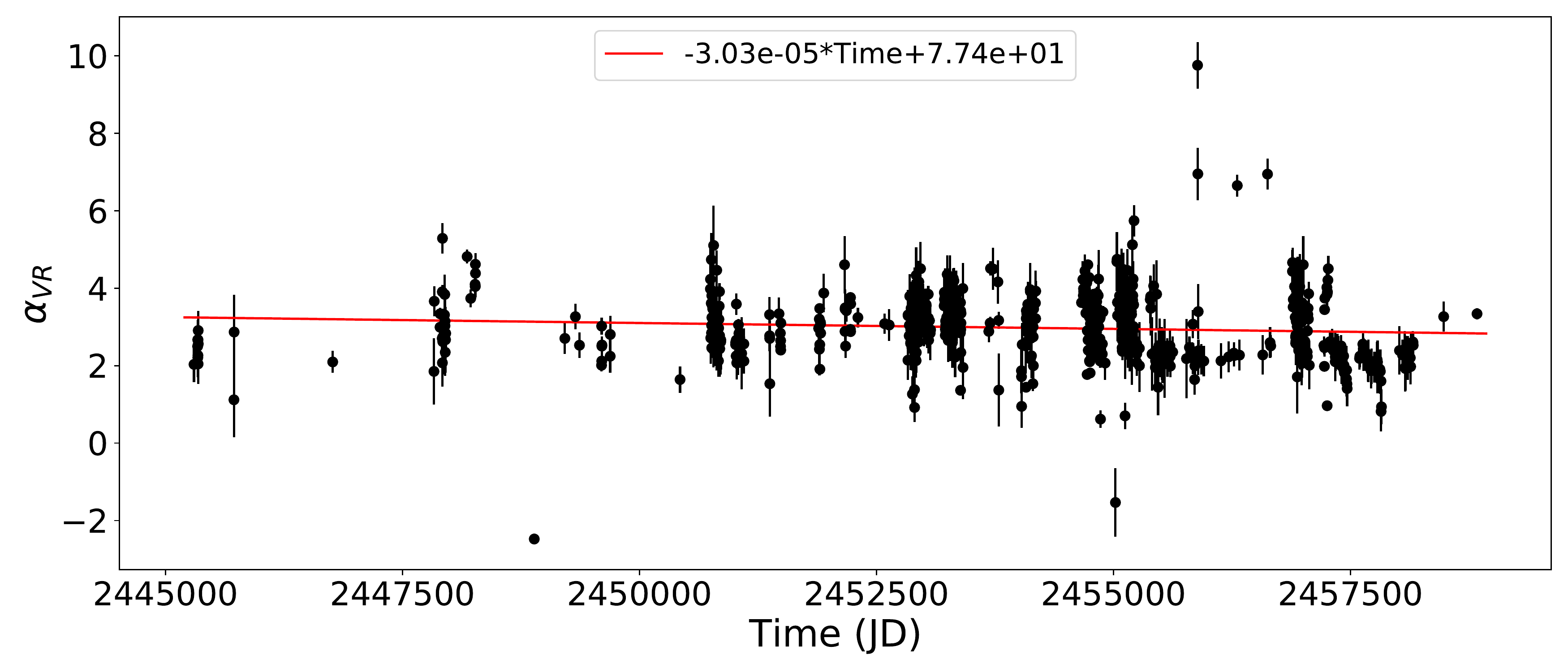}{0.5\textwidth}{(b)}}
    \caption{(a) Variation of spectral index ($\alpha_{VR}$) with $R$-band magnitude. (b) Variation of $\alpha_{VR}$ with time. The red line at each panel represents the linear fit.}
    \label{fig:specind}
\end{figure*}

\begin{deluxetable}{lrrrr}
\tablenum{2}
\tablecaption{Color variation with time in optical $U\!BV\!RI$ long-term lightcurves of AO\,0235+164 \label{tab:color_t}}
\tablewidth{0pt}
\tablehead{
\colhead{CI} & \colhead{$m$} &  \colhead{$c$} & \colhead{$\rho$} & \colhead{$p$}
}
\startdata
$U-B$ & $-$1.52E-05 &  3.74E+01 & $-$2.06E-01 &  8.28E-02  \\
$B-V$ & 6.58E-06  & $-$1.52E+01 & 1.42E-01  & 4.79E-03  \\
$V-R$ & $-$5.34E-06 &  1.38E+01 & $-$9.19E-02 &  1.39E-02  \\
$R-I$ & 1.83E-05  & $-$4.40E+01 & 2.85E-01  & 1.74E-08  \\
$U-I$ & 5.63E-05  & $-$1.35E+02 & 4.03E-01  & 1.88E-03  \\
$B-I$ & 4.16E-05  & $-$9.92E+01 & 4.50E-01  & 3.41E-11  \\
\enddata
\tablecomments{In the column headings: CI: color indices; $m$ = slope; $c$ = intercept; $\rho$ = Pearson coefficient; $p$ = null hypothesis probability for \autoref{fig:color_var}a}
\end{deluxetable}

\begin{deluxetable}{lrrrr}
\tablenum{3}
\tablecaption{Color variation with $R$-band magnitude in optical $U\!BV\!RI$ long-term lightcurves of AO\,0235+164 \label{tab:color_r}}
\tablewidth{0pt}
\tablehead{
\colhead{CI} & \colhead{$m$} &  \colhead{$c$} & \colhead{$\rho$} & \colhead{$p$}
}
\startdata
$U-B$ & $-$1.36E-01 &  2.37E+00 &  $-$5.37E-01 & 3.35E-05 \\
$B-V$ & 1.62E-02  & 7.04E-01  & 1.41E-01 & 7.41E-03 \\
$V-R$ & $-$3.54E-03 &  7.98E-01 &  $-$2.58E-02 & 4.92E-01 \\
$R-I$ & 1.62E-02  & 7.00E-01  & 1.37E-01 & 7.59E-03 \\
$U-I$ & $-$6.47E-02 &  3.85E+00 &  $-$2.07E-01 & 1.30E-01 \\
$B-I$ & 6.23E-02  & 1.66E+00  & 3.66E-01 & 1.69E-07 \\
\enddata
\tablecomments{In the column headings: CI: color indices; $m$ = slope; $c$ = intercept; $\rho$ = Pearson coefficient; $p$ = null hypothesis probability for \autoref{fig:color_var}b}
\end{deluxetable}

\noindent
The term \lq color' denotes the magnitude difference between two quasi-simultaneous observations at two different wavebands. We plotted the variation of optical colors ($U-B$, $B-V$, $V-R$, $R-I$, and $B-I$) with time and $R$-magnitude in \autoref{fig:color_var}. We listed the results of a straight line ($Y = mX + c$) fitting to all these plots in \autoref{tab:color_t} and \autoref{tab:color_r}. The linear fits of the color versus time plots do not show any trend, except for the  rather sparsely sampled ($B-I$) color, which has a high slope (4.16$\times$10$^{-5}$) in \autoref{fig:color_var}a, along with the highest Pearson correlation coefficient (0.45), and the lowest null hypothesis probability (3.41$\times$10$^{-11}$). Among the color versus magnitude relations, the strongest relationship is between ($B-I$) and $R$ (\autoref{fig:color_var}b), having a positive slope (6.23$\times$10$^{-2}$) with the highest Pearson coefficient (0.37) and the lowest $p$-value (1.69$\times$10$^{-7}$) (\autoref{tab:color_r}), indicates a bluer-when-brighter (BWB) trend when the widest range of the available colors is considered.

\begin{deluxetable}{lcccc}
\tablenum{4}
\tablecaption{Spetral index variation with $R$-band magnitude and time in optical $U\!BV\!RI$ long-term lightcurves of AO\,0235+164 \label{tab:specind_rt}}
\tablewidth{0pt}
\tablehead{
\colhead{Dependency} & \colhead{$m$} &  \colhead{$c$} & \colhead{$\rho$} & \colhead{$p$}
}
\startdata
$\alpha_{VR}$ vs $R$ & $-$2.01E-02 & 3.30E+00 & $-$2.58E-02 & 4.92E-01 \\
$\alpha_{VR}$ vs JD & $-$3.03E-05 & 7.74E+01 & $-$9.19E-02 & 1.39E-02 \\
\enddata
\tablecomments{In the column headings: $m$ = slope; $c$ = intercept; $\rho$ = Pearson coefficient; $p$ = null hypothesis probability for \autoref{fig:specind}.}
\end{deluxetable}

\begin{deluxetable}{lccc}
\tablenum{5}
\tablecaption{Equivalence between internal field star numbering in the CASLEO/CAHA data used in the IDV analyses and field-star numbering in other standard star charts during different observation seasons \label{tab:star_chart}}
\tablewidth{0pt}
\tablehead{
\colhead{Season} & \colhead{CASLEO/CAHA} &  \colhead{Heidelberg$^a$} & \colhead{GKM2001$^b$}
}
\startdata
1999--2001 & \phantom{1}2 & \phantom{1}8 & 10 \\
(CASLEO)  & \phantom{1}4 & C1	 & \phantom{1}9 \\
          & \phantom{1}5 & \phantom{1}6	 & 11 \\
          & \phantom{1}7 & --	 & \phantom{1}1 \\
          & \phantom{1}8 & --	 & \phantom{1}3 \\
          & 10 & --	 & \phantom{1}8 \\
          & 12 & --	 & 16 \\
\hline
2004--2005  & 2	& \phantom{1}8 & 10 \\
(CASLEO)    & 4	& C1 & \phantom{1}9 \\
            & 5	& \phantom{1}6 & 11 \\
            & 6	& -- & \phantom{1}8 \\
            & 7	& -- & \phantom{1}7 \\
\hline
2005        & \phantom{1}2 & 8 & 10 \\
(CAHA)      & 11 & C1 & \phantom{1}9 \\
            & 12 & -- & \phantom{1}1 \\
            & 13 & -- & \phantom{1}3 \\
            & 14 & -- & \phantom{1}7 \\
            & 15 & -- & \phantom{1}8 \\
            & 16 & 6 & 11 \\
            & 17 & -- & 16 \\
\hline
2018--2019 & 2 & \phantom{1}8 & 10 \\
(CASLEO)   & 4 & C1 & \phantom{1}9 \\
           & 5 & \phantom{1}6 & 11 \\
           & 6 & -- & \phantom{1}8 \\
           & 7 & -- & \phantom{1}7 \\
           & 8 & -- & 16 \\
\enddata
\tablecomments{$a.$ \url{https://www.lsw.uni-heidelberg.de/projects/extragalactic/charts/0235+164.html}\\
$b.$ \citet{GKM01}}
\end{deluxetable}

\begin{deluxetable*}{lc|cc|cccccc|c}
\tablenum{6}
\tablecaption{Result of scaled C-criterion and F-test for IDV on AO\,0235+164 differential lightcurves from CASLEO and CAHA 
\label{tab:idv_res1a}}
\tablewidth{0pt}
\tablehead{
\colhead{Date} & \colhead{JD} & \colhead{Band} &  \colhead{No. of} & \colhead{S1, S2} & \colhead{$\Gamma$} & \colhead{$C_{\Gamma}$} & \colhead{$F_{\Gamma}$} & \colhead{$F^{0.005}_c$} & \colhead{Status} & \colhead{Final}\\
 & \colhead{} & \colhead{} & \colhead{obs.} & \colhead{} & \colhead{} & \colhead{} & \colhead{} & \colhead{} & \colhead{} & \colhead{Status}
}
\startdata
    1999 Nov 2 & 2451485 & $V$ & 23 & 2,3 & 0.8886 & 11.3640 & 129.1405 & 3.1246 &  V & V \\
    & & & & 2,6 & 1.0867 & 12.9184 & 166.8856 & 3.1912 &  V & \\
    & & & & 2,10 & 1.6876 & 8.1627 & 66.6298 & 3.1246 &  V & \\
    & & & & 2,11 & 0.7431 & 13.4002 & 179.5650 & 3.1246 &  V & \\
     \hline
    1999 Nov 3 & 2451486 & $V$ & 22 & 2,3 & 1.0707 & 5.6976 & 32.4624 & 3.1347 &  V & V\\
    & & & & 2,11 & 0.8841 & 6.0726 & 36.8768 & 3.1347 &  V & \\
    \hline
    1999 Nov 4 & 2451487 & $R$ & 30 & 2,3 & 1.0059 & 8.4058 & 70.6582 & 2.6737 &  V & V \\
    & & & & 2,11 & 0.6639 & 9.8857 & 97.7278 & 2.6737 &  V & \\
    \cmidrule(lr){3-11}
    & & $V$ & 30 & 2,3 & 0.9994 & 8.9281 & 79.7104 & 2.6737 &  V & V\\
    & & & & 2,11 & 0.8286 & 9.6683 & 93.4770 & 2.6737 &  V & \\
    \hline
    1999 Nov 5 & 2451488 & $R$ & 23 & 2,3 & 1.4994 & 1.5631 & 2.4433 & 3.1246 &  NV & NV\\
    & & & & 2,11 & 0.9852 & 1.9303 & 3.7260 & 3.1246 &  NV & \\
    \cmidrule(lr){3-11}
    & & $V$ & 22 & 2,3 & 1.4403 & 3.0342 & 9.2064 & 3.1347 &  V & V \\
    \hline
    1999 Nov 6 & 2451489 & $R$ & 30 & 2,3 & 0.8471 & 17.5775 & 308.9682 & 2.6737 &  V & V \\
    & & & & 2,6 & 0.9769 & 12.3281 & 151.9824 & 2.6737 &  V & \\
    & & & & 2,7 & 1.3573 & 9.9373 & 98.7501 & 2.7048 &  V & \\
    & & & & 2,8 & 1.3805 & 9.8381 & 96.7876 & 2.7048 &  V & \\
    & & & & 2,10 & 1.6936 & 6.8657 & 47.1376 & 2.6737 &  V & \\
    & & & & 2,11 & 0.5616 & 15.4338 & 238.2019 & 2.6737 &  V & \\
    \cmidrule(lr){3-11}
    & & $V$ & 29 & 2,3 & 0.8485 & 18.1892 & 330.8486 & 2.7233 &  V & V \\
    & & & & 2,6 & 1.0013 & 11.7527 & 138.1254 & 2.7233 &  V & \\
    & & & & 2,7 & 1.3527 & 12.5480 & 157.4516 & 2.7397 &  V & \\
    & & & & 2,8 & 1.4133 & 13.4172 & 180.0214 & 2.7397 &  V & \\
    & & & & 2,10 & 1.5626 & 17.6674 & 312.1376 & 2.7233 &  V & \\
    & & & & 2,11 & 0.7018 & 17.9948 & 323.8145 & 2.7233 &  V & \\
    \hline
    1999 Nov 7 & 2451490 & $R$ & 11 & 2,3 & 0.9562 & 3.5930 & 12.9095 & 5.8479 &  V & PV \\
    & & & & 2,4 & 1.9798 & 2.2801 & 5.1990 & 5.8479 &  NV & \\
    & & & & 2,6 & 1.1143 & 4.3903 & 19.2751 & 5.8479 &  V & \\
    & & & & 2,10 & 1.9703 & 1.7073 & 2.9148 & 5.8479 &  NV & \\
    & & & & 2,11 & 0.6197 & 2.9496 & 8.7003 & 5.8479 &  V & \\
    \cmidrule(lr){3-11}
    & & $V$ & 12 & 2,3 & 0.9382 & 2.9304 & 8.5871 & 5.3191 &  V & PV \\
    & & & & 2,4 & 1.7807 & 1.9342 & 3.7410 & 5.3191 &  NV & \\
    & & & & 2,6 & 1.1169 & 2.8931 & 8.3701 & 5.3191 &  V & \\
    & & & & 2,10 & 1.7653 & 2.1046 & 4.4292 & 5.3191 &  NV & \\
    & & & & 2,11 & 0.7772 & 4.3359 & 18.7997 & 5.3191 &  V & \\
\enddata
\tablecomments{S1 and S2 are the comparison and control star numbers, respectively, used for the IDV tests. Star numbers follow the star maps shown in \autoref{tab:star_chart}.}
\end{deluxetable*}
\begin{deluxetable*}{lc|cc|cccccc|c}
\tablenum{6}
\tablecaption{Result of scaled C-test and F-test for IDV on AO\,0235+164 differential lightcurves from CASLEO and CAHA (continued...)\label{tab:idv_res1b}}
\tablewidth{0pt}
\tablehead{
\colhead{Date} & \colhead{JD} & \colhead{Band} &  \colhead{No. of} & \colhead{S1, S2} & \colhead{$\Gamma$} & \colhead{$C_{\Gamma}$} & \colhead{$F_{\Gamma}$} & \colhead{$F^{0.005}_c$} & \colhead{Status} & \colhead{Final} \\
 & \colhead{} & \colhead{} & \colhead{obs.} & \colhead{} & \colhead{} & \colhead{} & \colhead{} & \colhead{} & \colhead{} & \colhead{status}}

\startdata
    2000 Dec 21 & 2451900 & $R$ & 10 & 2,3 & 0.9446 & 2.3638 & 5.5876 & 6.5402 &  NV & PV \\
    & & & & 2,6 & 1.0793 & 4.9877 & 24.8767 & 6.5402 &  V & \\
    & & & & 2,7 & 1.5020 & 2.0187 & 4.0753 & 6.5402 &  NV & \\
    & & & & 2,8 & 1.5289 & 1.9985 & 3.9939 & 6.5402 &  NV & \\
    & & & & 2,9 & 0.8790 & 2.5120 & 6.3100 & 6.5402 &  NV & \\
    & & & & 2,11 & 0.6246 & 7.4168 & 55.0085 & 6.5402 &  V & \\
    \cmidrule(lr){3-11}
    & & $V$ & 10 & 2,3 & 0.9509 & 3.4671 & 12.0208 & 6.5402 &  V & PV \\
    & & & & 2,6 & 1.1202 & 2.4789 & 6.1449 & 6.5402 &  NV & \\
    & & & & 2,7 & 1.5357 & 1.8729 & 3.5079 & 6.5402 &  NV & \\
    & & & & 2,8 & 1.6064 & 2.0031 & 4.0124 & 6.5402 &  NV & \\
    & & & & 2,9 & 1.0966 & 3.7299 & 13.9120 & 6.5402 &  V & \\
    & & & & 2,11 & 0.7842 & 1.5920 & 2.5343 & 6.5402 &  NV & \\
    \hline
    2000 Dec 23 & 2451902 & $R$ & 10 & 2,3 & 0.8588 & 4.4475 & 19.7803 & 6.5402 &  V & V \\
    & & & & 2,6 & 0.9890 & 5.1629 & 26.6559 & 6.5402 &  V & \\
    & & & & 2,7 & 1.3855 & 3.5919 & 12.9020 & 6.5402 &  V & \\
    & & & & 2,8 & 1.4091 & 2.8222 & 7.9646 & 6.5402 &  V & \\
    & & & & 2,9 & 0.8000 & 4.6739 & 21.8451 & 6.5402 &  V & \\
    & & & & 2,11 & 0.5664 & 5.3690 & 28.8267 & 6.5402 &  V & \\
    & & & & 2,13 & 1.7083 & 3.0181 & 9.1089 & 6.5402 &  V & \\
    \cmidrule(lr){3-11}
    & & $V$ & 11 & 2,3 & 0.8509 & 6.5241 & 42.5634 & 5.8479 &  V & PV \\
    & & & & 2,6 & 1.0031 & 5.4277 & 29.4602 & 5.8479 &  V & \\
    & & & & 2,7 & 1.3714 & 5.0139 & 25.1395 & 5.8479 &  V & \\
    & & & & 2,8 & 1.4341 & 5.2879 & 27.9619 & 5.8479 &  V & \\
    & & & & 2,9 & 0.9797 & 1.4805 & 2.1919 & 5.8479 &  NV & \\
    & & & & 2,11 & 0.7013 & 5.1765 & 26.7965 & 5.8479 &  V & \\
    & & & & 2,13 & 1.5668 & 4.3770 & 19.1586 & 5.8479 &  V & \\
\enddata
\tablecomments{S1 and S2 are the comparison and control star numbers respectively used for the IDV tests. Star numbers follow the star maps shown in \autoref{tab:star_chart}.}
\end{deluxetable*}
\begin{deluxetable*}{lc|cc|cccccc|c}
\tablenum{6}
\tablecaption{Result of scaled C-test and F-test for IDV on AO\,0235+164 differential lightcurves from CASLEO and CAHA (continued...)\label{tab:idv_res1c}}
\tablewidth{0pt}
\tablehead{
\colhead{Date} & \colhead{JD} & \colhead{Band} &  \colhead{No. of} & \colhead{S1, S2} & \colhead{$\Gamma$} & \colhead{$C_{\Gamma}$} & \colhead{$F_{\Gamma}$} & \colhead{$F^{0.005}_c$} & \colhead{Status} & \colhead{Final} \\
 & \colhead{} & \colhead{} & \colhead{obs.} & \colhead{} & \colhead{} & \colhead{} & \colhead{} & \colhead{} & \colhead{} & \colhead{status}}
\startdata
    2001 Nov 9 & 2452223 & $R$ & 12 & 2,11 & 1.2042 & 4.6476 & 21.5998 & 5.3191 &  V & V \\
    \cmidrule(lr){3-11}
    & & $V$ & 12 & 2,3 & 1.8778 & 2.5035 & 6.2675 & 5.3191 &  NV & NV \\
    & & & & 2,4 & 3.6191 & 1.1450 & 1.3111 & 5.3191 &  NV & \\
    & & & & 2,9 & 2.2039 & 1.2380 & 1.5326 & 5.4171 &  NV & \\
    & & & & 2,10 & 3.5871 & 2.0056 & 4.0226 & 5.3191 &  NV & \\
    & & & & 2,11 & 1.5366 & 1.7566 & 3.0857 & 5.3191 &  NV & \\
    \hline
    2001 Nov 10 & 2452224 & $R$ & 10 & 2,3 & 2.3728 & 1.0570 & 1.1172 & 6.5402 &  NV & NV \\
    & & & & 2,9 & 2.2429 & 1.1058 & 1.2229 & 6.5402 &  NV & \\
    & & & & 2,11 & 1.5595 & 0.9395 & 0.8826 & 6.5402 &  NV & \\
    \cmidrule(lr){3-11}
    & & $V$ & 10 & 2,3 & 2.3876 & 1.0788 & 1.1637 & 6.5402 &  NV & NV \\
    & & & & 2,9 & 2.7847 & 1.3860 & 1.9209 & 6.5402 &  NV & \\
    & & & & 2,11 & 1.9713 & 0.9038 & 0.8168 & 6.5402 &  NV & \\
    \hline
    2001 Nov 11 & 2452225 & $R$ & 14 & 2,3 & 2.0291 & 1.4125 & 1.9951 & 4.5724 &  NV & NV \\
    & & & & 2,9 & 1.5447 & 1.2505 & 1.5638 & 4.6425 &  NV & \\
    & & & & 2,11 & 1.3171 & 1.6860 & 2.8427 & 4.5724 &  NV & \\
    \cmidrule(lr){3-11}
    & & $V$ & 14 & 2,3 & 2.0291 & 1.4125 & 1.9951 & 4.5724 &  NV & NV \\
    & & & & 2,9 & 1.5447 & 1.2505 & 1.5638 & 4.6425 &  NV & \\
    & & & & 2,11 & 1.3171 & 1.6860 & 2.8427 & 4.5724 &  NV & \\
    \hline
    2001 Nov 12 & 2452226 & $R$ & 12 & 2,3 & 1.8479 & 1.5819 & 2.5025 & 5.3191 &  NV & PV \\
    & & & & 2,11 & 1.2074 & 3.0203 & 9.1222 & 5.3191 &  V & \\
    \cmidrule(lr){3-11}
    & & $V$ & 12 & 2,3 & 1.8704 & 1.9230 & 3.6980 & 5.3191 &  NV & NV \\
    & & & & 2,4 & 3.5981 & 1.0281 & 1.0571 & 5.3191 &  NV & \\
    & & & & 2,10 & 3.5672 & 2.3374 & 5.4634 & 5.3191 &  NV & \\
    & & & & 2,11 & 1.5330 & 1.5642 & 2.4468 & 5.3191 &  NV & \\
    \hline
    2001 Nov 13 & 2452227 & $R$ & 11 & 3,4 & 2.0213 & 1.1434 & 1.3073 & 5.8479 &  NV & NV \\
    \cmidrule(lr){3-11}
    & & $V$ & 11 & 3,4 & 1.1840 & 0.6858 & 0.4703 & 5.8479 &  NV & NV\\
\enddata
\tablecomments{S1 and S2 are the comparison and control star numbers respectively used for the IDV tests. Star numbers follow the star maps shown in \autoref{tab:star_chart}.}
\end{deluxetable*}
\begin{deluxetable*}{lc|cc|cccccc|c}
\tablenum{6}
\tablecaption{Result of scaled C-test and F-test for IDV on AO\,0235+164 differential lightcurves from CASLEO and CAHA (continued...)\label{tab:idv_res1d}}
\tablewidth{0pt}
\tablehead{
\colhead{Date} & \colhead{JD} & \colhead{Band} &  \colhead{No. of} & \colhead{S1, S2} & \colhead{$\Gamma$} & \colhead{$C_{\Gamma}$} & \colhead{$F_{\Gamma}$} & \colhead{$F^{0.005}_c$} & \colhead{Status} & \colhead{Final} \\
 & \colhead{} & \colhead{} & \colhead{obs.} & \colhead{} & \colhead{} & \colhead{} & \colhead{} & \colhead{} & \colhead{} & \colhead{status}}
\startdata
    2005 Jan 16 & 2453387 & $R$ & 11 & 2,3 & 1.5238 & 3.8074 & 14.4962 & 5.8479 &  V & V \\
    & & & & 2,4 & 3.3465 & 2.7388 & 7.5010 & 5.8479 &  V & \\
    & & & & 2,6 & 3.3316 & 2.7442 & 7.5308 & 5.8479 &  V & \\
    & & & & 2,7 & 1.4051 & 3.4058 & 11.5996 & 5.8479 &  V & \\
    \hline
    2005 Nov 2 & 2453677 & $R$ & 32 & 2,3 & 1.2848 & 6.4237 & 41.2636 & 2.5846 &  V & V \\
    & & & & 2,4 & 0.8959 & 4.5227 & 20.4545 & 2.5846 &  V & \\
    & & & & 2,5 & 0.2571 & 4.3013 & 18.5013 & 2.5846 &  V & \\
    & & & & 2,6 & 0.5453 & 3.9310 & 15.4528 & 2.5846 &  V & \\
    & & & & 2,7 & 0.5844 & 4.9283 & 24.2884 & 2.5846 &  V & \\
    & & & & 2,8 & 0.4738 & 7.0560 & 49.7865 & 2.5846 &  V & \\
    & & & & 2,9 & 0.3415 & 6.5374 & 42.7373 & 2.5846 &  V & \\
    & & & & 2,10 & 0.3397 & 4.0828 & 16.6695 & 2.5846 &  V & \\
    \hline
    2005 Nov 4 & 2453679 & $R$ & 12 & 2,3 & 0.8534 & 4.3059 & 18.5409 & 5.3191 &  V & V \\
    & & & & 2,4 & 0.5599 & 3.7978 & 14.4235 & 5.3191 &  V & \\
    & & & & 2,5 & 0.1421 & 5.0805 & 25.8111 & 5.3191 &  V & \\
    & & & & 2,6 & 0.3029 & 5.3341 & 28.4524 & 5.3191 &  V & \\
    & & & & 2,7 & 0.3534 & 7.6846 & 59.0525 & 5.3191 &  V & \\
    & & & & 2,8 & 0.2839 & 4.3153 & 18.6220 & 5.3191 &  V & \\
    & & & & 2,9 & 0.1914 & 11.0664 & 122.4647 & 5.3191 &  V & \\
    & & & & 2,10 & 0.1875 & 5.4019 & 29.1804 & 5.3191 &  V & \\
    \hline
    2005 Nov 5 & 2453680 & $R$ & 44 & 2,3 & 0.9749 & 10.6766 & 113.9894 & 2.2266 &  V & V \\
    & & & & 2,4 & 0.6398 & 9.3431 & 87.2939 & 2.2266 &  V & \\
    & & & & 2,5 & 0.1942 & 11.0439 & 121.9674 & 2.2266 &  V & \\
    & & & & 2,6 & 0.3721 & 10.7338 & 115.2142 & 2.2266 &  V & \\
    & & & & 2,7 & 0.4059 & 10.2100 & 104.2433 & 2.2266 &  V & \\
    & & & & 2,8 & 0.3427 & 8.3494 & 69.7127 & 2.2266 &  V & \\
    & & & & 2,9 & 0.2399 & 12.1459 & 147.5239 & 2.2266 &  V & \\
    & & & & 2,10 & 0.2340 & 8.5775 & 73.5744 & 2.2341 &  V & \\
    \hline
    2005 Nov 6 & 2453681 & $R$ & 40 & 2,3 & 1.0022 & 7.8517 & 61.6495 & 2.3212 &  V & V \\
    & & & & 2,4 & 0.6946 & 8.8524 & 78.3645 & 2.3212 &  V & \\
    & & & & 2,5 & 0.2051 & 6.6830 & 44.6620 & 2.3212 &  V & \\
    & & & & 2,6 & 0.4022 & 7.6630 & 58.7211 & 2.3212 &  V & \\
    & & & & 2,8 & 0.3694 & 5.9489 & 35.3890 & 2.3212 &  V & \\
    & & & & 2,9 & 0.2576 & 6.8684 & 47.1751 & 2.3212 &  V & \\
    & & & & 2,10 & 0.2563 & 5.1520 & 26.5433 & 2.3212 &  V & \\
\enddata
\tablecomments{S1 and S2 are the comparison and control star numbers respectively used for the IDV tests. Star numbers follow the star maps shown in \autoref{tab:star_chart}.}
\end{deluxetable*}


\begin{deluxetable*}{lc|cc|cccccc|c}
\tablenum{6}
\tablecaption{Result of scaled C-test and F-test for IDV on AO\,0235+164 differential lightcurves from CASLEO and CAHA (continued...)\label{tab:idv_res1e}}
\tablewidth{0pt}
\tablehead{
\colhead{Date} & \colhead{JD} & \colhead{Band} &  \colhead{No. of} & \colhead{S1, S2} & \colhead{$\Gamma$} & \colhead{$C_{\Gamma}$} & \colhead{$F_{\Gamma}$} & \colhead{$F^{0.005}_c$} & \colhead{Status} & \colhead{Final} \\
& \colhead{} & \colhead{} & \colhead{obs.} & \colhead{} & \colhead{} & \colhead{} & \colhead{} & \colhead{} & \colhead{} & \colhead{status}}
\startdata
    2005 Nov 8 & 2453683 & $R$ & 28 & 2,3 & 0.9329 & 2.4256 & 5.8834 & 2.7940 &  NV & NV \\
    & & & & 2,4 & 0.6336 & 2.3363 & 5.4585 & 2.7770 &  NV & \\
    & & & & 2,5 & 0.1788 & 1.7843 & 3.1836 & 2.7770 &  NV & \\
    & & & & 2,6 & 0.3598 & 2.2163 & 4.9120 & 2.7770 &  NV & \\
    & & & & 2,7 & 0.4059 & 1.4945 & 2.2335 & 2.7770 &  NV & \\
    & & & & 2,8 & 0.3451 & 2.1606 & 4.6682 & 2.9002 &  NV & \\
    & & & & 2,9 & 0.2318 & 1.3895 & 1.9307 & 2.7770 &  NV & \\
    \hline
    2005 Dec 5 & 2453710 & $R$ & 20 & 2,3 & 1.4796 & 1.4053 & 1.9748 & 3.4317 &  NV & NV \\
    & & & & 2,4 & 1.0247 & 0.7240 & 0.5242 & 3.4317 &  NV & \\
    & & & & 2,5 & 0.3133 & 0.9355 & 0.8752 & 3.4317 &  NV & \\
    & & & & 2,6 & 0.6030 & 1.1896 & 1.4151 & 3.4317 &  NV & \\
    & & & & 2,8 & 0.5634 & 1.0332 & 1.0674 & 3.4317 &  NV & \\
    & & & & 2,9 & 0.3979 & 1.0716 & 1.1482 & 3.4317 &  NV & \\
    & & & & 2,10 & 0.3915 & 0.8994 & 0.8089 & 3.4317 &  NV & \\
    \hline
    2005 Dec 6 & 2453711 & $R$ & 16 & 2,3 & 1.4092 & 2.1709 & 4.7129 & 4.0698 &  NV & PV \\
    & & & & 2,4 & 0.9785 & 3.7432 & 14.0118 & 4.0698 &  V & \\
    & & & & 2,5 & 0.2848 & 2.1323 & 4.5467 & 4.0698 &  NV & \\
    & & & & 2,6 & 0.5570 & 2.7562 & 7.5967 & 4.0698 &  V & \\
    & & & & 2,7 & 0.6157 & 1.8489 & 3.4186 & 4.0698 &  NV & \\
    & & & & 2,8 & 0.5266 & 1.3717 & 1.8815 & 4.0698 &  NV & \\
    & & & & 2,8 & 0.5266 & 1.3717 & 1.8815 & 4.0698 &  NV & \\
    & & & & 2,9 & 0.3691 & 2.4056 & 5.7869 & 4.0698 &  NV & \\
    & & & & 2,10 & 0.3688 & 2.0671 & 4.2727 & 4.0698 &  NV & \\
    \hline
    2019 Dec 17 & 2458835 & $R$ & 30 & 9,10 & 1.3151 & 1.2773 & 1.6315 & 2.6740 &  NV & NV \\
    & & & & 9,11 & 0.7377 & 1.3155 & 1.7307 & 2.6740 &  NV & \\
    & & & & 9,12 & 1.0425 & 1.0698 & 1.1445 & 2.6740 &  NV & \\
\enddata
\tablecomments{S1 and S2 are the comparison and control star numbers respectively used for the IDV tests. Star numbers follow the star maps shown in \autoref{tab:star_chart}.}
\end{deluxetable*}


\begin{deluxetable*}{lcc|cccc|ccc|ccc}
\tablenum{7}
\tablecaption{Result of power enhanced F-test and nested ANOVA test for IDV on AO\,0235+164 differential lightcurves from CASLEO and CAHA\label{tab:idv_res2}}
\tablewidth{0pt}
\tablehead{
\colhead{Obs.} & \colhead{Band} &  \colhead{No. of} &
\multicolumn4c{Power enhanced F-test} & \multicolumn3c{Nested ANOVA test} & \colhead{Status} & \colhead{Variability} & \colhead{doubling} \\
\cmidrule(lr){4-7} \cmidrule(lr){8-10}
\colhead{date} & \colhead{} & \colhead{Obs.} & \colhead{Comp.}  & \colhead{} & \colhead{} & \colhead{} & \colhead{} & \colhead{} & \colhead{} & \colhead{} & \colhead{amplitude(\%)} & \colhead{timescale}\\ 
\colhead{} & \colhead{} & \colhead{} & \colhead{star} & \colhead{DOF($\nu_1$,$\nu_2$)} & \colhead{$F_{\text{enh}}$} & \colhead{$F^{0.005}_c$} & \colhead{DOF($\nu_1$,$\nu_2$)} & \colhead{$F$} & \colhead{$F^{0.005}_c$} & \colhead{} & \colhead{} & \colhead{(days)}}
\startdata
        1999 Nov 2 & $V$ & 23 & 2 & (22, 87) & 116.132 & 2.209 & (5, 17) & 58.924 & 5.075 & V  & 43.99 & 0.103 \\
        1999 Nov 3 & $V$ & 22 & 2 & (21, 42)  & 34.529 & 2.540 & (5, 16) & 10.920 & 5.212 & V & 24.47 & 0.145 \\
        1999 Nov 4 & $V$ & 30 & 2 & (29, 58) & 86.046 & 2.216 & (7, 22) & 38.922 & 4.109 & V & 34.48 & 0.106 \\ 
                    & $R$ & 30 &   & (29, 58) & 82.016 & 2.216 & (7, 22) & 40.356 & 4.109 & V & 32.59 & 0.083 \\ 
        1999 Nov 5 & $V$ & 22 & 2 & (21, 21) & 9.207 & 3.216 & (5, 16) & 4.426  & 5.212 & NV & 10.94 & 0.140 \\
                    & $R$ & 23 &   & (22, 44) & 2.951 & 2.487 & (5, 17) & 9.426  & 5.075 & V & 9.03 & 0.335 \\
        1999 Nov 6 & $V$ & 29 & 2 & (28, 166) & 211.363 & 1.960 & (7, 21) & 58.114 & 4.179 & V  & 36.79 & 0.092 \\
                    & $R$ & 30 &   & (29, 170) & 107.913 & 1.941 & (7, 22) & 74.686  & 4.109 & V  & 37.90 & 0.085 \\
        1999 Nov 7 & $V$ & 12 & 2 & (11, 55) & 6.392 & 2.854  & -- & -- & -- & PV  & 9.13 & 0.170 \\
                    & $R$ & 11 &   & (10, 50) & 6.413 & 2.988  & -- & -- & -- & PV  & 5.36 & 0.244 \\
        2000 Dec 21 & $V$ & 10 & 2 & (9, 54) & 4.813 & 3.055  & -- & -- & -- & PV  & 6.95 & 0.275 \\
                    & $R$ & 10 &   & (9, 54) & 6.73 & 3.055  & -- & -- & -- & PV  & 7.67 & 0.428 \\
        2000 Dec 23 & $V$ & 11 & 2 & (10, 70) & 10.314 & 2.846  & -- & -- & -- & PV  & 20.58 & 0.200 \\
                    & $R$ & 10 &   & (9, 63) & 14.542 & 2.989  & -- & -- & -- & PV  & 14.18 & 0.180 \\
        2001 Nov 9 & $V$ & 12 & 2 & (11, 54) & 2.345 & 2.863  & -- & -- & -- & NV  & 12.13 & 0.372 \\
                    & $R$ & 12 &   & (11, 22) & 5.91 & 3.612  & -- & -- & -- & PV  & 12.73 & 0.441 \\
        2001 Nov 10 & $V$ & 10 & 2 & (9, 27) & 1.152 & 3.557  & -- & -- & -- & NV  & 8.49 & 0.227 \\
                    & $R$ & 10 &   & (9, 27) & 1.054 & 3.557  & -- & -- & -- & NV  & 5.64 & 0.660 \\
        2001 Nov 11 & $V$ & 14 & 2 & (13, 38) & 2.02 & 2.923  & -- & -- & -- & NV  & 9.63 & 0.364 \\
                    & $R$ & 14 &    & (13, 38) & 2.02 & 2.923  & -- & -- & -- & NV & 9.63 & 0.364 \\
        2001 Nov 12 & $V$ & 12 & 2 & (11, 44) & 2.212 & 2.969  & -- & -- & -- & NV  & 12.06 & 0.539 \\
                &    $R$ & 12 & 2 & (11, 22) & 3.928 & 3.612  & -- & -- & -- & PV  & 11.74 & 0.856 \\
        2001 Nov 13 & $V$ & 11 & 3 & (10, 10) & 0.470 & 5.847  & -- & -- & -- & NV  & 10.81 & 0.160 \\
                &    $R$ & 11 & 3 & (10, 10) & 1.307 & 5.847  & -- & -- & -- & NV  & 10.19 & 0.178 \\
        2005 Jan 16 & $R$ & 11 & 2 & (10, 40) & 9.842 & 3.117  & -- & -- & -- & PV  & 32.92 & 0.095\\
        2005 Nov 2 & $R$ & 32 & 2 & (31, 247) & 27.709 & 1.868  & (7, 24) & 37.156 & 3.991 & V  & 8.98 & 0.189 \\
        2005 Nov 4 & $R$ & 12 & 2 & (11, 88) & 31.995 & 2.689  & -- & -- & -- & V & 6.59 & 0.166 \\
        2005 Nov 5 & $R$ & 44 & 2 & (43, 343) & 124.459 & 1.713  & (10, 33) & 16.301 & 3.26 & V  & 13.60 & 0.146 \\
        2005 Nov 6 & $R$ & 40 & 2 & (39, 273) & 57.755 & 1.767  & (9, 30) & 87.95 & 3.45 & V  & 9.79 & 0.227 \\
        2005 Nov 8 & $R$ & 28 & 2 & (27, 182) & 4.371 & 1.965  & (6, 21) & 0.449 & 4.393 & PV  & 3.18 & 0.365 \\
        2005 Dec 5 & $R$ & 20 & 2 & (19, 133) & 1.067 & 2.200  & (4, 15) & 14.394 & 5.803 & PV  & 2.61 & 0.391 \\
        2005 Dec 6 & $R$ & 16 & 2 & (15, 120) & 4.863 & 2.373  & -- & -- & -- & PV  & 3.53 & 0.746 \\ 
        2019 Dec 17 & $R$ & 30 & 9 & (29, 87) & 1.453 & 2.075  & (7, 22) & 2.341 & 4.109 & NV  & 7.74 & 0.038 \\ 
\enddata
\tablecomments{Comparison star numbers follow the star maps shown in \autoref{tab:star_chart}.}
\end{deluxetable*}
\subsubsection{Spectral Variations and SEDs} \label{sec:spec}

\noindent
We plotted the optical ($BV\!R$) spectral energy distributions for the nights where observations were taken at all of these three filters. Following the prescription of \citet{2005A&A...438...39R}, we took into account the total absorption by the Milky Way galaxy and the foreground absorber at $z = 0.524$, and subtracted the extinction magnitudes ($A_U=2.519$, $A_B=1.904$, $A_V=1.473$, $A_R=1.260$, $A_I=0.902$)
from the calibrated magnitudes of respective wavebands and then converted them into extinction-corrected flux densities, $F_{\nu}$. The accompanying video contains one-day averaged optical SEDs for those 360 nights (An example frame is shown in \autoref{fig:SEDmovie}). \autoref{fig:spec} shows a few examples of SEDs of low, moderate, and high flux states, plotted in ($\nu F_{\nu}$ -- $\nu$) format. Mostly, the SEDs have a declining shape following a power law. However, there are evidences of spectral hardening on several nights (e.g., JD 2445337, JD 2445721, JD 2448889, JD 2452901, JD 2453230). \\
\\
From the one-day binned multiwavelength lightcurves we calculated the spectral indices ($\alpha_{VR}$) for all the days when the source was observed in both $V$ and $R$ bands, using the formula given by \citet{2015A&A...573A..69W} on extinction corrected magnitudes, as

\begin{equation}
    \alpha_{VR} = \frac{0.4 (V-R)}{\log(\nu_V/\nu_R)}\, ,
\end{equation}
where $\nu_V$ and $\nu_R$ respectively represent the effective frequencies of $V$ and $R$ band filters \citep{2005ARAA43293B}. We plotted the variation of spectral indices with time and $R$-band magnitude (\autoref{fig:specind}) and listed the results of linear fits, Pearson coefficient, and null hypothesis probability in \autoref{tab:specind_rt}. We do not find any significant long-term variation of the spectral index with time, nor is there a correlation with $R$-magnitude.

\subsection{Intraday Variability} \label{sec:idv}

\begin{figure*}
    \centering
    \includegraphics[width=\textwidth]{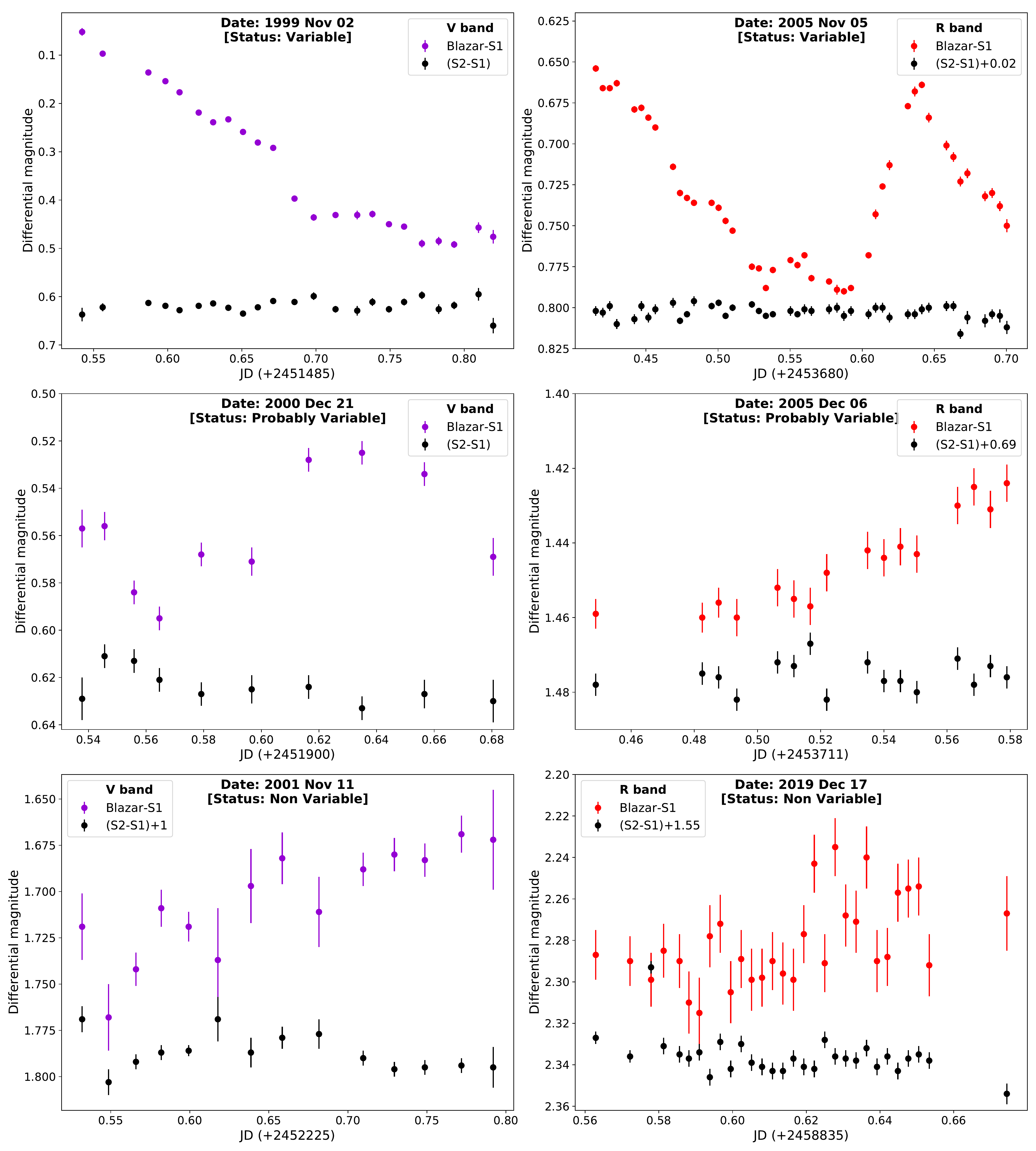}
    \caption{Some intraday lightcurves of AO\,0235+164 on nights when the source showed different states of variability. S1 and S2 represent the comparison and control star respectively. In some panels, the differential lightcurve of the control star is shifted to bring it into the same frame of the blazar DLC for better visual comparison of variability.}
    \label{fig:idv_lc}
\end{figure*}

\noindent
We applied four frequently used statistical tests for IDV: scaled $C$-criterion, scaled $F$-test, the power-enhanced $F$-test, and the nested analysis of variance (ANOVA) test \citep{2014AJ....148...93D, 2015AJ....150...44D, ZAC-2017, ZAC-2020} to detect statistically significant intraday flux variability in AO\,0235+164 lightcurves observed by CASLEO and CAHA telescopes. These tests mainly compare the variations in blazar magnitudes with the variations in magnitudes of one or more stars within the field-of-view of the blazar and have different advantages and disadvantages. We collected data from multiple field stars along with the blazar data (\autoref{tab:star_chart}). We applied the first three methods on the intraday differential lightcurves of AO\,0235+164 where at least 10 observations were recorded per night with at least one optical filter between 1999 November 2 to 2019 December 17. We employed the nested ANOVA test only on lightcurves having at least 20 observations per night.

\begin{figure}
	\centering
	 \includegraphics[width=0.5\textwidth]{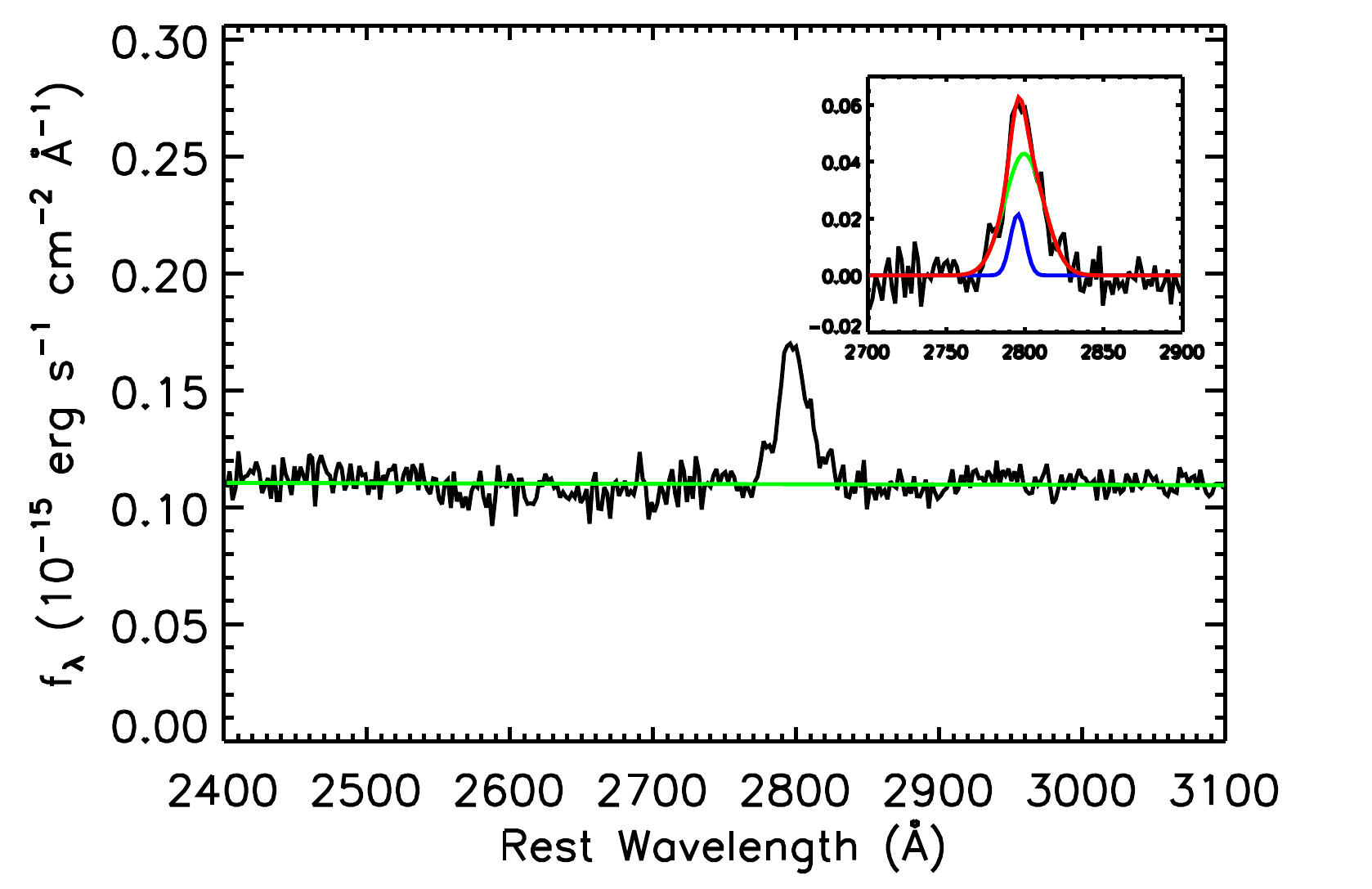}
	 \caption{Spectral fitting of AO\,0235+164, where the black line is the original spectrum while the green line is the single power law for the fitted continuum. The inset shows Mg II line fitting  where the blue, green, and red lines are the narrow, broad, and total components, respectively.}
	\label{fig:BHM}
\end{figure}

\subsubsection{Scaled C-criterion} \label{sec:ctest}

\noindent
Differential photometry, where the blazar magnitudes are compared to one or more stars in the same field of view, is the usual technique for obtaining blazar lightcurves free from the effects of any non-astrophysical fluctuations. The simplest differential photometry involves a single comparison star, while a second star, whose magnitudes are measured against the same comparison star, is used for a stability check. We denote B, S1, and S2 as the blazar, comparison, and control star, respectively. The variability test requires two differential lightcurves (DLC): (blazar–comparison star) and (control star–comparison star). The latter is believed to be affected only by instrumental fluctuations as any known or suspected variable star can be discarded.\\ 
\\
\citet{1997AJ....114..565J} and \citet{1999A&AS..135..477R} introduced a parameter $C$ defined as $C = \sigma_{\rm B-S1}/ \sigma_{\rm S2-S1}$, where $\sigma_{\rm B-S1}$ and $\sigma_{\rm S2-S1}$ are the standard deviations in blazar DLC and control star DLC, respectively. The blazar is considered to be variable with 99.5 per cent confidence level if $C$ is greater than a critical value of 2.576. \\
\\
\citet{HMW88} pointed out that it is important to select non-variable stars with magnitudes close to the blazar magnitude as comparison and control stars. Otherwise, even if the blazar is non-variable, there will be difference between $\sigma_{\rm B-S1}$ and $\sigma_{\rm S2-S1}$ due to differences in photon statistics and other random-noise terms (sky, read-out noise). To use field stars with different magnitude levels, \citet{HMW88} suggests calculating a correction factor $\Gamma$ to scale $\sigma_{\rm S2-S1}$ to the instrumental level of $\sigma_{\rm B-S1}$ for proper comparison. $\Gamma$ can be estimated using the following formula:

\begin{equation} \label{eqn:hwg}
    \Gamma^2 = \left(\frac{N_{\rm S2}}{N_B}\right)^2 \left[\frac{N_{\rm S1}^2(N_B+P)+N_B^2(N_{\rm S1}+P)}{N_{\rm S2}^2(N_{\rm S1}+P)+N_{\rm S1}^2(N_{\rm S2}+P)}\right]
\end{equation}
where $N$ is the total (sky-subtracted) counts within the aperture, while the sub-indices B, S1 and S2 correspond to $N$ of the blazar, comparison star and control star, respectively. The factor $P$ contains the common noise-terms, as $P = n_{\rm pix} (N_{\rm sky} + N^2_{\rm RON})$, where $n_{\rm pix}$ is the number of pixels within the aperture, $N_{\rm sky}$ is the sky level and $N_{\rm RON}$ is the read-out noise. We used the median values of $N$ of the objects and sky for calculating $\Gamma$. Thus, the scaled $C$ parameter ($C_{\Gamma}$) is defined as

\begin{equation}
    C_{\Gamma} = \frac{C}{\Gamma} = \frac{1}{\Gamma}\left(\frac{\sigma_{\rm B-S1}}{ \sigma_{\rm S2-S1}}\right).
\end{equation}
The source is considered variable if $C_{\Gamma} \geq 2.576$. Even though the C parameter is not a proper statistic, it remains a useful indicator of stability \citep{2014AJ....148...93D, 2015AJ....150...44D, ZAC-2017, ZAC-2020}.

\subsubsection{Scaled F-test} \label{sec:ftest}
\noindent
The standard F-statistics parameter is  $F = \sigma_{\rm B-S1}^2/ \sigma_{\rm S2-S1}^2$, where $\sigma_{\rm B-S1}^2$ and $\sigma_{\rm S2-S1}^2$ are the variances in blazar DLC and a control star DLC respectively. The scaled F-statistics $F_{\Gamma}$ is given as

\begin{equation*}
    F_{\Gamma} = \frac{F}{\Gamma^2} = \frac{1}{\Gamma^2}\left(\frac{\sigma_{\rm B-S1}^2}{ \sigma_{\rm S2-S1}^2}\right).
\end{equation*}
The F-statistic assumes that the uncertainties in the observations are normally distributed. If $n_{\rm (B-S1)}$ and $n_{\rm (S2-S1)}$ are the sizes of the blazar and control star DLC respectively, the number of degrees of freedom in the numerator and denominator of the F-statistic are $\nu_1 = n_{\rm (B-S1)}-1$ and $\nu_2 = n_{\rm (S2-S1)}-1$, respectively. We calculated $F_{\Gamma}$ and considered the blazar to be variable with 99.5 per cent confidence if $F_{\Gamma}$ was greater than the critical value $F_c^{\alpha} (\nu_1, \nu_2)$ at $\alpha=0.005$ \citep{ZAC-2017, ZAC-2020}.

\subsubsection{Power-enhanced F-test} \label{sec:peftest}
\noindent
The power-enhanced F -test (PEF) has been used in various recent blazar IDV studies \citep[and references therein]{2019ApJ...871..192P, Pandey2020}. The power-enhanced F-statistic has the advantage of comparing the blazar variance to the combined variance of multiple field stars and is given as \citep{2014AJ....148...93D}

\begin{equation}
    F_{\rm enh} = \frac{s_{\rm blz}^2}{s_c^2},
\end{equation}
where $s_{\rm blz}^2$ is the variance of the DLC of the blazar with respect to a reference star, and $s_c^2$ is the combined variance of the comparison stars' DLCs with respect to the reference star. Thus, $s_c^2$ is given as

\begin{equation}
    s_c^2 = \frac{1}{\left(\sum_{j=1}^k n_j\right)-k} \sum_{j=1}^k \sum_{i=1}^{n_j} s_{j,i}^2.
\end{equation}
Here, $k$ is the total number of available comparison stars in the DLC, $n_j$ is the number of observations of the $j^{\rm th}$ comparison star, and $s_{j,i}^2$ is the scaled square deviation of the $i^{\rm th}$ observation of the $j^{\rm th}$ comparison star given as

\begin{equation}
    s_{j,i}^2 = \Gamma_j (m_{j,i}-\Bar{m_j})^2.
\end{equation}
Here $\Gamma_j$ is the scale factor of the $j^{\rm th}$ comparison star DLC computed following \autoref{eqn:hwg}. \\
\\
Using the data of the field stars, we first checked the star--star DLCs to identify any spikes due to instrumental errors or improper removal of cosmic rays, and removed them iteratively if they were more than 3 standard deviations from the mean magnitude. We considered a \lq\lq well-behaved" star with low fluctuations and an average magnitude close to the blazar as the reference star. The number of degrees of freedom in the numerator and denominator of the F-statistics are $\nu_1 = n_{\rm blz}-1$ and $\nu_2 = \left(\sum_{j=1}^k n_j\right)-k$, respectively. We calculated $F_{\rm enh}$, and considered the blazar to be variable (V) with 99.5 percent confidence if $F_{\rm enh}$ was greater than the critical value $F_c (\nu_1, \nu_2)$ at $\alpha=0.005$.

\subsubsection{Nested ANOVA test} \label{sec:nanova}

\noindent
In the nested analysis of variance (ANOVA) test, DLCs of the blazar are generated with respect to all the comparison stars used as reference stars. The details of this method are given in \citet{2015AJ....150...44D}. The nested ANOVA test needs a large number of points in the light curves, strongly limiting its application to densely populated DLCs. We divided the DLCs with at least 20 observations into groups such that each group contains 4 observations. Equation (4) of \citet{2015AJ....150...44D} considers an ideal set of lightcurves where the total number of observations are divisible by the group size. In most of the DLCs in this work, the total number of observations was not an integral multiple of the group size of 4. So, in those cases, the last group contained less than 4 observations, and we calculated the degrees of freedom accordingly to compute the mean square due to groups (MS$_G$) and mean square due to the nested observations in groups (MS$_{O(G)}$). The ANOVA F-statistic is given as, $F = \text{MS}_G/\text{MS}_{O(G)}$. For a significance level of $\alpha=0.005$, if the F -statistic is greater than the critical value ($F_c$), the blazar is taken as variable (V), otherwise as non-variable (NV) with 99.5 per cent confidence. \\
\\
We have listed the results of the scaled C-criterion and scaled F-test in \autoref{tab:idv_res1a} and those of power enhanced F-test and the nested ANOVA test in \autoref{tab:idv_res2}. In the case of scaled C-criterion and F-test, we fixed one particular star as the comparison star for each dataset. The source is declared variable with respect to one comparison-control star pair if both scaled C-statistics and F-statistics cross their respective critical values. We declare the final variability status of the blazar as variable/non-variable (V/NV) if it is variable/non-variable against all control stars. If the blazar is variable against some of the control stars, we call it probably variable (PV). We did not carry out the nested ANOVA test in a few datasets containing less than 20 observations. In the case of the power-enhanced F-test in absence of the corresponding nested ANOVA test, we call the blazar probably variable (PV) even if the F-statistic crosses the critical value, as the F-test is more prone to give a false positive result \citep{ZAC-2017,ZAC-2020}. If nested ANOVA is present and both the tests cross the critical values, we call the blazar variable (V). Otherwise, we declare the source non-variable (NV). We list the summary of the IDV tests in \autoref{tab:idv_sum}. We give a final verdict on the variability status of the source after comparing the results of the combination of the C-test and F-test (C\&F) from \autoref{tab:idv_res1a} and results of the combination of the power-enhance F-test and nested ANOVA test (P\&N) from \autoref{tab:idv_res2}. If the results from both combinations were the same, we kept that result. If C\&F declared \lq\lq V" and P\&N declared \lq\lq PV" due to the absence of nested ANOVA, we finally consider the source variable (V). We considered variability on 2005 November 8 as \lq\lq NV" because both C-test and nested ANOVA resulted in non-variability. Despite being variable in nested ANOVA, we consider the 2005 December 5 lightcurve \lq\lq NV" as the F-test and PEF-test detected no variability. A few examples of DLCs of AO\,0235+164 having different variability characteristics (V/PV/NV) are shown in \autoref{fig:idv_lc}.

\subsubsection{Doubling timescale} \label{sec:dt}

\noindent
A flux doubling/halving timescale gives an estimate of the variability timescale ($\tau_\mathrm{var}$) of a source. We calculate the flux doubling/halving timescale ($\tau_d$) between two consecutive observations and its corresponding significance ($\sigma$) as

\begin{equation}
\begin{split}
    \mathcal{F}(t_{i+1}) = \mathcal{F}(t_i)* 2^{(t_{i+1}-t_i)/\tau_d} \\
    \sigma = |\mathcal{F}(t_{i+1})-\mathcal{F}(t_{i})|/\varepsilon_i ,
\end{split}
\end{equation}
where $\mathcal{F}(t_i)$ and $\varepsilon_i$ are the flux observed at time $t_i$ and the corresponding measurement uncertainty, respectively. We consider the fastest doubling timescale ($\tau_d^\mathrm{min}$) with a higher significance than $3\sigma$ as an estimate for $\tau_\mathrm{var}$. We obtained $\tau_d^\mathrm{min} < 1$ day for all the nights when the source showed significant IDV both in scaled F-test and nested ANOVA test. This further strengthens our claims for the  frequent presence of IDV. Following \autoref{eq:vamp} we computed the variability amplitudes on the same nights. All these results are listed in \autoref{tab:idv_res2}.

\begin{deluxetable}{lc|cc|c}
\tablenum{8}
\tablecaption{Summary of statistical tests for IDV on AO\,0235+164 differential lightcurves from CASLEO and CAHA\label{tab:idv_sum}}
\tablewidth{0pt}
\tablehead{
\colhead{Obs.} & \colhead{Band} & \multicolumn2c{Combined variability status} & \colhead{Final} \\
\cmidrule(lr){3-4}
\colhead{date} & \colhead{} & \colhead{($C$ \& $F$-test)$^a$} & \colhead{(PEF \&} & \colhead{status} \\
\colhead{} & \colhead{} & \colhead{} & \colhead{N-ANOVA)$^b$} & \colhead{}}
\startdata
    1999 Nov 2  & $V$ &  V  &  V   &  V \\
    1999 Nov 3  & $V$ &  V  &  V   &  V \\
    1999 Nov 4  & $V$ &  V  &  V   &  V \\
                & $R$ &  V  &  V   &  V \\
    1999 Nov 5  & $V$ &  V  &  NV  & PV \\
                & $R$ &  NV &  V   & PV \\
    1999 Nov 6  & $V$ &  V  &  V   &  V \\
                & $R$ &  V  &  V   &  V \\
    1999 Nov 7  & $V$ &  PV &  PV  & PV \\
                & $R$ &  PV &  PV  & PV \\
    2000 Dec 21 & $V$ &  PV &  PV  & PV \\
                & $R$ &  PV &  PV  & PV \\
    2000 Dec 23 & $V$ &  PV &  PV  & PV \\
                & $R$ &  V  &  PV  & V  \\
    2001 Nov 9  & $V$ &  NV &  NV  & NV \\
                & $R$ &  V  &  PV  & V  \\
    2001 Nov 10 & $V$ &  NV &  NV  & NV \\
                & $R$ &  NV &  NV  & NV \\
    2001 Nov 11 & $V$ &  NV &  NV  & NV \\
                & $R$ &  NV &  NV  & NV \\
    2001 Nov 12 & $V$ &  NV &  NV  & NV \\
                & $R$ &  PV &  PV  & PV \\
    2001 Nov 13 & $V$ &  NV &  NV  & NV \\
                & $R$ &  NV &  NV  & NV \\
    2005 Jan 16 & $R$ &  V  &  PV  & V  \\
    2005 Nov 2  & $R$ &  V  &  V   & V  \\
    2005 Nov 4  & $R$ &  V  &  V   & V  \\
    2005 Nov 5  & $R$ &  V  &  V   & V  \\
    2005 Nov 6  & $R$ &  V  &  V   & V  \\
    2005 Nov 8  & $R$ &  NV &  PV  & NV \\
    2005 Dec 5  & $R$ &  NV &  PV  & NV \\
    2005 Dec 6  & $R$ &  PV &  PV  & PV \\
    2019 Dec 17 & $R$ &  NV &  NV  & NV \\
\enddata
\tablecomments{$^a$\autoref{tab:idv_res1a}, $^b$\autoref{tab:idv_res2}, PEF=power-enhanced F-test.}
\end{deluxetable}

\subsubsection{Duty cycle}
\label{sec:duty}

We calculated the duty cycle (DC) of AO\,0235+164 using the definition of \citet{1999A&AS..135..477R}, that was used later by multiple authors \citep[e.g.,][]{stalin2009, 2016MNRAS.455..680A}. The formula for DC for a particular waveband is given as,

\begin{equation}
    \text{DC} = 100 \frac{\sum_{i=1}^n N_i (1/\Delta t_i)}{\sum_{i=1}^n (1/\Delta t_i)} \% 
\end{equation}
where $\Delta t_i=\Delta t_{i,obs}/(1+z)$ (duration of the monitoring session on $i^{\text{th}}$ night is $\Delta t_{i,obs}$). Thus, this formula calculates the duty cycle weighted by the cosmological redshift corrected monitoring duration of each night. We set $N_i=\text{1, 0.5, and 0}$ for the nights with variability status \lq\lq V", \lq\lq PV", and \lq\lq NV" respectively. We obtained the duty cycle of AO\,0235+164 to be $\sim$44 percent in $V$-band, and $\sim$45 percent in $R$-band considering the nights where the source was observed for at least 2 hours.
\subsection{The mass of the central black hole}
\label{sec:BH}

\noindent
We estimate the mass of the SMBH in AO\,0235+164 by using its spectrum observed using the CCD Imaging/Spectropolarimeter (SPOL) at the Steward Observatory\footnote{\url{http://james.as.arizona.edu/~psmith/Fermi}} on 2011 January 8 (air mass = 1.12). This spectrum was selected since the blazar was then at its lowest level during the period 2008--2018, and should ensure the best visibility of the emission lines because of the lower continuum contribution from the jet. The observed wavelength range of the spectrum we used is 4000--7550 \AA, with a spectral resolution of 4 \AA, and it is analyzed by following the procedure given in \citet{2020MNRAS.491...92L}. Firstly, it was corrected for Galactic extinction with the reddening map of \citet{1998ApJ...500..525S}, and then was shifted to the rest-frame wavelength by using the redshift of 0.94.\\
\\
This spectral coverage meant we could use the Mg II line, which is prominent on the spectrum shown in \autoref{fig:BHM} (focused on the 2400$-$3100 \AA\ range), to estimate the SMBH mass. We modeled the continuum by applying a single power law ($f_{\lambda} \propto \lambda^{\alpha}$) (as Fe II emission is rather weak). A Gaussian profile was then used to fit the Mg II line, centered at the position of 2800 \AA, on the continuum-subtracted spectrum. The broad component of Mg II was fitted with a Gaussian with a 1000\,$\rm km\,s^{-1}$ lower limit, while a Gaussian with upper limit of 1000\,$\rm km\,s^{-1}$ was applied for the narrow component. In order to estimate the corresponding errors of full width at half maximum (FWHM) and flux, we generated 100 mock spectra by adding random Gaussian noise to the original spectrum using the flux density errors, and then took the standard deviation of measurements from those mock spectra as the uncertainties. Here, the flux density errors were the RMS value of the spectrum calculated over the spectral window of (3000$-$3100) \AA, after subtracting a second-order polynomial function. \autoref{fig:BHM} shows the resulting fit to the spectrum. Our best fitting results indicate that the line width of the broad Mg II component is FWHM =  3151 $\rm km\,s^{-1}$, with log-scale luminosity in $\rm erg~s^{-1}$,  $\log(L_{\rm Mg II})$ = 42.8. \\
\\
The line width and the  Mg II line luminosity we find are consistent with the range of values FWHM=3100--3500 $\rm km\,s^{-1}$ and $\log(L_{\rm Mg II})$=42.5--42.8, respectively, which were derived by \citet{raiteri2007} from one VLT and four TNG spectra of AO 0235+164 acquired in 2003--2004. We use the FWHM and luminosity of the broad Mg II line, not the continuum luminosity, as we are unable to  exclude the jet emission contribution, despite the low state spectrum that we could use for this blazar. The black hole mass is derived from the empirical relation used for Mg II \citep{2006ChJAA...6..396K}, which is based on measured broad line region sizes in the reverberation-mapping AGN sample of \citet{2004ApJ...613..682P}, as

\begin{equation}
\frac{M_{\rm BH}}{\rm M_{\odot}}=2.9\times10^{6} \left(\frac{L_{\rm Mg II}}{10^{42}\
\rm erg\ s^{-1}} \right)^{0.57\pm0.12} \left(\frac{\rm FWHM_{\rm Mg II}}{10^{3}\ \rm km\ s^{-1}} \right)^{2}
\end{equation}
Thus, the SMBH mass is $\log(M_{\rm BH}/M_{\odot}) = 7.90 \pm 0.25 $,  where the uncertainty is estimated from the measurement uncertainties of the FWHM and luminosity of Mg II. Using optical spectroscopy data from the SDSS archive, \citet{2021ApJS..253...46P} reported a somewhat higher mass, $\log(M_{\rm BH}/M_{\odot})$ = 8.58 $\pm$ 0.34, and an accretion disk luminosity (in erg s$^{-1}$), of $\log(L_{\rm disk})$ = 45.30 $\pm$ 0.22. Using the method mentioned in \citet{2021ApJS..253...46P} with $\log(L_{\rm Mg II})$ = 42.8, we obtained a lower disk luminosity (in erg s$^{-1}$) of $\log(L_{\rm disk})$ = 45.01 $\pm$ 0.20 from the spectrum observed on 2011 January 8.

\section{Discussion} \label{sec:disc}

\begin{figure}
    \centering
    \includegraphics[width=0.45\textwidth]{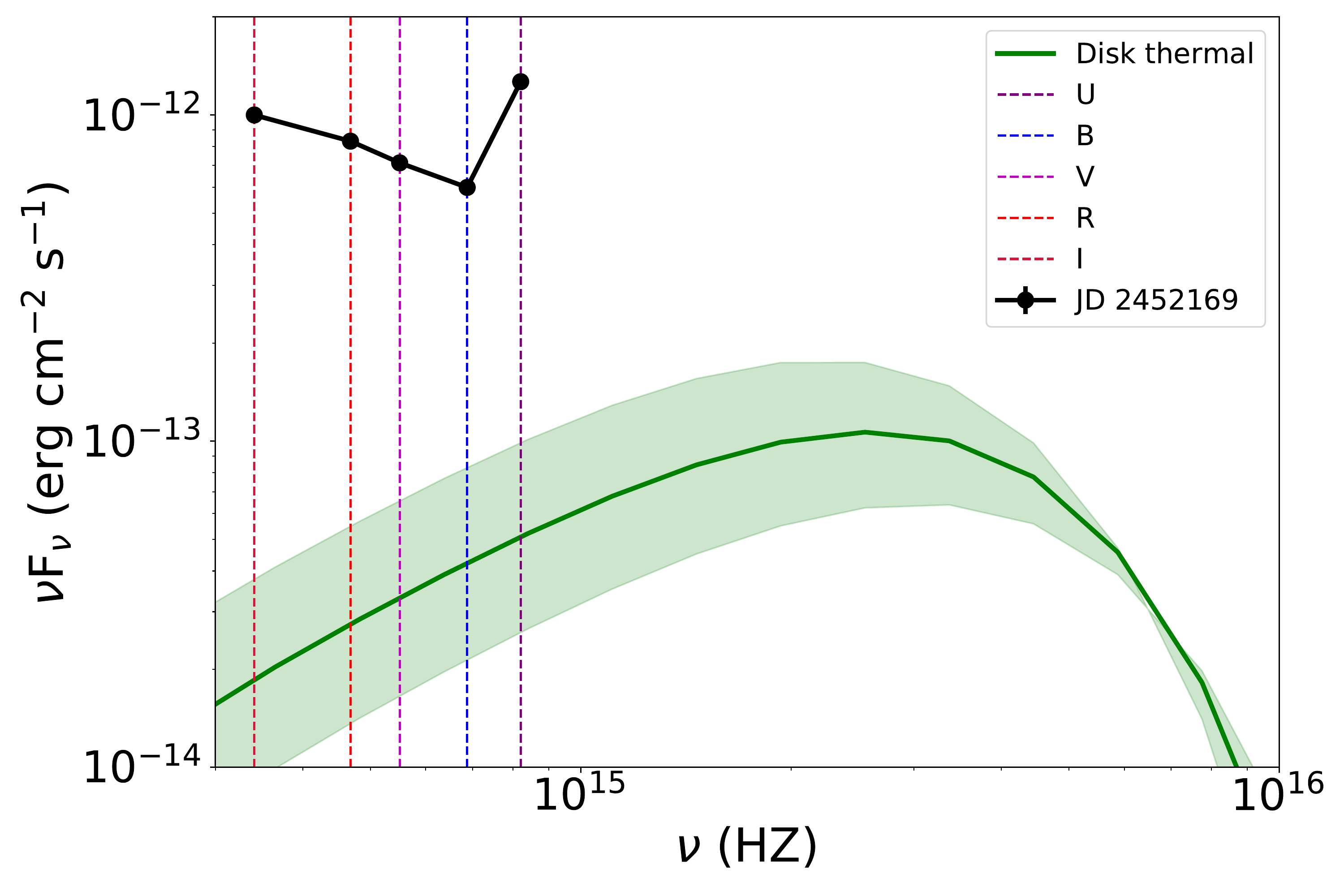}
    \caption{Comparison of the SED of the lowest flux state observed on JD 2452169 and the thermal emission from the accretion disk in the observer's frame. The thermal emission component is calculated using a multi-temperature disk model with the black hole mass $\log(M_{\rm BH}/M_{\odot})$ = 7.9$\pm$0.25, and the log-scale disk luminosity in erg s$^{-1}$,  $\log(L_{\rm disk})$ = 45.01$\pm$0.20. The shaded region indicates the uncertainties in the calculation of the disk thermal component.}
    \label{fig:disk_comp}
\end{figure}

\noindent
In this work, we have presented a detailed temporal and spectral study of the  highly variable emission from the blazar AO\,0235+164 observed at multiple optical wavebands ($U\!BV\!RI$) from October 1975 to December 2019. The lightcurves have highly uneven data sampling due to gaps in observation seasons and non-uniform observation campaigns. Although $U$-band data are quite sparsely sampled the $BV\!RI$ observations have denser sampling when the source was highly active. Multiple long-term studies suggested that AO\,0235+164 shows $\sim$2-year long flaring episodes with multiple sub-flares after intervals of $\sim$8 years \citep{2006A&A...459..731R, 2017ApJ...837...45F,2022MNRAS.513.5238R}. \autoref{fig:mwlc} shows a difference of about six magnitudes between the quiescent and outburst states in all optical wavebands, corresponding to an energy flux variation of more than two orders of magnitude (\autoref{fig:spec}). The long-term variability amplitudes at all five wavebands are quite similar (\autoref{tab:ltvres}). Also, we found a strong correlation with zero time-lag between the $U\!BV\!I$ observations and the $R$-band data (\autoref{fig:corr} and \autoref{fig:dcf}), which implies a common radiative process at a single emission zone is responsible for the bulk of the emission at the optical wavebands. \\
\\
Sometimes during the quiescent states of powerful blazars, the disk thermal emission component becomes  visible as a big blue bump on top of the synchrotron emission component from the jet in the optical-UV wavebands \citep[e.g.,][]{2021MNRAS.504.1103R}. As the disk emission is bluer than the jet synchrotron emission, an increase in the jet activity during low flux states displays a redder-when-brighter trend. The enhanced jet activity is observed when the charged particles inside the jet get accelerated to higher energies, and then radiate faster. Thus, the jet synchrotron component tends to get bluer with the increase in flux. If the jet emission completely outshines the disk emission, we expect to see a bluer-when-brighter trend \citep[e.g.,][]{Isler2017}. The flux increment can also be attributed to the increase in the jet Doppler factor \citep[e.g.,][]{2007A&A...470..857P}, which blueshifts the spectrum and produces a bluer-when-brighter trend because of the convexity of the spectrum. Such a trend is seen in the $(B-I)$ vs $R$ magnitude diagram (\autoref{fig:color_var}b) and indicates the domination of non-thermal jet emission over the thermal emission component of the accretion disk during both flaring and quiescent states. From the convex shapes of the optical $BV\!R$ SEDs during states ranging from quiescent to flaring (see the accompanying SED video and \autoref{fig:spec}), we may infer that the effect of the disk thermal emission is not significant in optical wavebands even during the low flux states. \\
\\
This can be explained in terms of the nature of disk thermal emission given the disk luminosity and the central black hole mass computed in \autoref{sec:BH}.
The primary, and most precise, black hole mass estimation methods are based on stellar and gas kinematics and reverberation mapping \citep[e.g.][]{2004ASPC..311...69V}. These methods need high spatial resolution spectroscopy data from the host galaxy and/or higher ionization emission lines and are not applicable to most BL Lacertae objects (BL Lacs). But in BL Lacs, if the weak emission lines are present, we can use the empirical methods \citep{2006ChJAA...6..396K} for BH mass estimation. The most common methods used for BH mass estimation for BL Lacs are the shortest variability timescales and periods of QPOs \citep{2012NewA...17....8G}. Since BL Lacs are highly variable objects, any BH mass estimation may be treated as an upper limit, and there are possibilities of detection of a shorter variability timescale or shorter QPO period. We obtained a log-scale BH mass of 7.90$\pm$0.25 in solar mass unit. The Steward observatory spectrum we used in our analysis had a narrower Mg II emission line (FWHM=3151 km s$^{-1}$) than those of \citet{raiteri2007} and \citet{2021ApJS..253...46P}, thus resulting in a lower mass estimate.
We considered a multi-temperature blackbody type accretion disk model, where the temperature at any portion of the disk is a function of the disk luminosity and the central black hole mass, to compute the thermal emission component. In \autoref{fig:disk_comp} we plotted the thermal component along with the optical-UV SED during the lowest activity state of AO\,0235+164 observed on JD 2452169. It is evident that, as the thermal emission peaks at far UV frequencies ($\sim$3.5$\times 10^{15}$ Hz) in the observer's frame of reference, the jet emission always dominates in $BV\!RI$ wavebands. We do not see any significant trend in the variation of the $(V-R)$ spectral index ($\alpha_{VR}$) (\autoref{fig:specind}).
The sudden rise of the $U$-band flux in \autoref{fig:disk_comp} is an indicator of a probable UV-soft X-ray bump as discussed in \citet{2005A&A...438...39R, 2006A&A...459..731R}. According to these studies, the source of the bump is either an additional synchrotron component coming from a separate emission region in the jet or the emission of a continuous inhomogeneous jet is suppressed in near UV region due to a discontinuity in opacity or misalignment of that particular emission region. \citet{2012ApJ...751..159A} mentioned that the whole optical-UV spectrum is produced by a single synchrotron emitting zone as the shape of the bump does not change with luminosity. They attributed the UV spectral hardening to an artifact due to the overestimation of extinction by \citet{2004ApJ...614..658J}.\\
For the detection of any statistically significant intraday variability in 33 lightcurves of AO\,0235+164 observed at CASLEO/CAHA, we employed different statistical tests widely used in AGN variability studies. The reliability of each of these tests has been disputed \citep[e.g.][]{2015AJ....150...44D, ZAC-2017}, so we here employed a comparative approach that could allow us to circumvent the limitations affecting any individual test. In the first place, we used the scaled $C$-criterion and the $F$-test. The first compares the dispersion of the blazar lightcurve to the dispersion of a field star (control star), while the latter does so with the variances. According  to  \citet{ZAC-2017} and \citet{ZAC-2020}, the F-test has a tendency to classify noisy non-variable curves as a variable (i.e., give false positives), while the $C$-criterion tends to give false negatives. Even though the $C$-criterion \citep{1999A&AS..135..477R} cannot be considered as an actual statistical test, it may still be a useful parameter to detect variability with high significance. The $F$-test, on the other hand, does not always work as expected, because it is particularly sensitive to non-Gaussian errors (``red noise''), which are usually an issue when analyzing blazars DLCs. \\
\\
We also used the power-enhanced F-test and the nested ANOVA test, which involve multiple field stars. It is expected that the power-enhanced F-test may also suffer from the same drawback of detecting false variability as the (original) $F$-test. In the nested ANOVA test, in turn, data grouping may lead to false results if data within a time span larger than the (unknown) variability timescale are grouped. Comparing the results of \autoref{tab:idv_res1a} and \autoref{tab:idv_res2}, while considering the tendencies of giving false results by the respective tests, we can confirm that the source was significantly variable in 4 out of 13 $V$-band lightcurves, and 9 out of 20 $R$-band lightcurves. The source seems to be probably variable in 3 $V$-band and 4 $R$-band lightcurves, and non-variable in the rest. On 1999 November 5, the combination of $C$-criterion and $F$-test indicates non-variability but the combination of power-enhanced $F$-test and nested ANOVA detects variability in the $R$-band lightcurve. The results in the $V$-band lightcurve on that day are exactly the opposite. Similar situations were observed also on 2001 November 9 and 2001 November 12. A visual inspection of the DLCs of these nights reveals that the blazar DLCs were classified as non-variable when either the control star DLC had higher variability (1999 November 5) or the measurement errors of the blazar DLCs were higher due to its low-flux state (2001 November 9 and 12). Higher measurement errors lead to a lower chance of significant variability detection. These strange results may be an example of the drawbacks of the applied methods when trying to recover low-amplitude variations from DLCs affected by non-Gaussian noise (part of the observations on that night were taken at air mass $>2$ and under non-photometric conditions). Otherwise, the combined results of different methods seem to more or less agree. Alongside the optical SED patterns, such frequent IDV establishes AO\,0235+164 as a low-energy peaked BL Lac (LBL) object. High energy peaked BL Lacs (HBL) show significantly less optical intraday variability than the LBLs \citep{1998A&A...329..853H, 1999A&AS..135..477R}.\\
\\
The differences in IDV behavior have been attributed to the strength of magnetic fields present in the jet of HBLs. A higher axial magnetic field ($B$) than a critical value ($B_c$) may prevent the generation of any bends and Kelvin-Helmhotz instabilities in the jet-base responsible for creating intraday microvariabilities. This indicates the presence of a weaker magnetic field than $B_c$ in the jet of AO\,0235+164. The critical magnetic field ($B_c$) is given in \citet{1995Ap&SS.234...49R} as

\begin{equation}
    B_c = \sqrt{4 \uppi n_e m_e c^2 (\Gamma^2-1)}/\Gamma,
\end{equation}
where $n_e$ is the electron density in the emission region, $m_e$ is the electron rest mass, and here $\Gamma$ is the bulk Lorentz factor of the jet flow. Considering a typical set of parameters, $n_e=429$ cm$^{-3}$ and $\Gamma=20$ \citep{ackermann2012}, we get $B_c \simeq 0.07$ G. \\

\begin{deluxetable}{ccc}
\tablenum{9}
\tablecaption{Variation of duty cycle with the duration of observation in $R$-band.}\label{tab:DC_var}
\tablewidth{0pt}
\tablehead{
\colhead{Observation} & \colhead{No. of} & \colhead{Duty}\\
\colhead{duration (hours)} & \colhead{nights} & \colhead{cycle (\%)}}
\startdata
    $>1$ & 20 & 52 \\
    $>2$ & 19 & 45 \\
    $>3$ & 17 & 50 \\
    $>4$ & 14 & 57 \\
    $>5$ & 13 & 64 \\
    $>6$ & 8  & 77 \\
\enddata  
\end{deluxetable}

From \autoref{tab:idv_res2} and \autoref{fig:idv_lc}, we can say that the variability amplitudes were higher in the 1999 season when the source was in a fainter state (higher magnitude) than its brighter state in the 2005 season. \citet{Marscher2013} suggested that enhancement of flux can arise from a more uniform flow of particles inside the jet, which in turn decreases the amplitude of microvariability associated with the turbulence inside the jet. \autoref{tab:DC_var} indicates that the probability of detection of significant variability increases with the duration of observation. Similar results for other blazars were found by \citet{2005A&A...440..855G}, \citet{rani2010}, and \citet{2016MNRAS.455..680A}.

From the flux doubling timescales listed in \autoref{tab:idv_res2}, we can estimate the upper limit to the size of the emission region ($R_{\rm max}$) using the light travel-time argument given as

\begin{equation}
    R_{\rm max} = \frac{c \delta t_{\rm var}}{1+z}
\end{equation}
where $z$ is the cosmological redshift of 0.94, $t_{\rm var}$ is the variability timescale, and $\delta$ is the Doppler boost of the jet. Considering $\delta = 24$ \citep{2009A&A...494..527H} and $t_{\rm var}$ to be the shortest flux doubling timescale of 0.083 days (when the source was significantly variable), we obtain an emission region size upper limit of $\sim 2.6 \times10^{15}$ cm. Assuming a conical jet model where the emission region fills up the entire jet cross-section, we can estimate the probable maximum distance ($d_{\rm max}$) of the emission region from the central black hole as, $d_{\rm max} = \Gamma R_{\rm max} = 5.2\times10^{16}$ cm.
To explain the observed strong variability, \citet{2016A&A...591A..21M} attempted to apply a swinging jet model that attributes the observed variability to a change in the viewing angle of the emission region with time (i.e. variation in the associated bulk Doppler factor). They reported a high rate of change in viewing angle of about 7$-$10 arcmin per day, considering a mean viewing angle of 2.3$^{\circ}$, would be necessary. However, they found that  this geometric wiggling-jet scenario was disfavored when considering the observed variation in color index with time.
Several earlier studies on AO 0235+164 associated the observed fast optical variability with  gravitational microlensing by the foreground absorber at $z=0.524$. \citet{2000AJ....120...41W} proposed that the 1997 flare resulted due to microlensing because of an observed correlation with zero lag between radio and optical lightcurves following \citet{1988A&A...198L..13S}, but the absence of any correlated flare in the X-ray lightcurve makes this explanation less likely. \citet{1993ApJ...415..101A} and \citet{raiteri2007} explained that such microlensing events can produce small amounts of fast flux amplification but are unlikely to dominate the high variability observed in AO 0235+164.

\section{conclusions} \label{sec:conc}

\noindent
In this work, we conducted a study of long-term and short-term (intraday) variability in the optical multiwaveband observations of the blazar AO\,0235+164. Here we summarize our results and the probable physical scenarios.\\
\\
\begin{enumerate}
    \item {We observed a variation of about six magnitudes between the quiescent and flaring episodes, or over two orders of magnitude variation in the SEDs.}
    \item {$U\!BV\!I$ lightcurves are highly correlated with the $R$-band lightcurve with zero time lag.}
    \item {A significant bluer-when-brighter trend is observed in the $(B-I)$ color variation with $R$-magnitude.}
    \item {All the optical $BV\!R$-band SEDs show convexity. These observations indicate that the optical emission is dominated by jet radiation.}
    \item {AO\, 0235+164 frequently shows statistically significant intraday variability in optical wavebands. This implies that AO\,0235+164 is an LBL and probably has a weak magnetic field in the jet environment.}
    \item {From the analysis of a broad Mg II emission line in a spectrum of AO\, 0235+164 taken at a low state, we estimate a central black-hole mass of $\sim 7.9\times 10^7 M_{\odot}$.}
\end{enumerate}

\acknowledgments
\noindent
Data from the Steward Observatory spectropolarimetric monitoring project were used. This program is supported by Fermi Guest Investigator grants NNX08AW56G, NNX09AU10G, NNX12AO93G, and NNX15AU81G. This paper has made use of up-to-date SMARTS optical/near-infrared light curves that are available at {\url{www.astro.yale.edu/smarts/glast/home.php}}. This work is partly based on data taken and assembled by the WEBT collaboration and stored in the WEBT archive at the Osservatorio Astrofisico di Torino - INAF (\url{https://www.oato.inaf.it/blazars/webt/}). These data are available upon request to the WEBT President Massimo Villata (\href{mailto:massimo.villata@inaf.it}{massimo.villata@inaf.it}). This work is based on data acquired at Complejo Astron\'omico El Leoncito, operated under an agreement between the Consejo Nacional de Investigaciones Cient\'ificas y T\'ecnicas de la Rep\'ublica Argentina and the National Universities of La Plata, C\'ordoba and San Juan. We thank Anabella Araudo and Ileana Andruchow for help with the observations made with CASLEO and the data analysis. \\
\\
We thankfully acknowledge the anonymous reviewer for very useful comments which helped us to improve the manuscript.
We acknowledge the support of the Department of Atomic Energy, Government of India, under project identification number RTI 4002. ACG is partially supported by Chinese Academy of Sciences (CAS) President’s International Fellowship Initiative (PIFI) (grant no. 2016VMB073). GER acknowledges support from grants PIP 0554 (CONICET), PIP 2021-1639 (CONICET), and grant PID2019-105510GBC31 of the Spanish Ministerio de Ciencia, Innovación y Universidades and through the Center of Excellence Mara de Maeztu 2020-2023 award to the ICCUB (CEX2019-000918-M). JAC is Mar\'ia Zambrano researcher fellow funded by the European Union -NextGenerationEU- (UJAR02MZ), supported by PIP 0113 (CONICET) and PICT-2017-2865 (ANPCyT). JAC was also supported by grant PID2019-105510GB-C32/AEI/10.13039/501100011033 from the Agencia Estatal de Investigaci\'on of the Spanish Ministerio de Ciencia, Innovaci\'on y Universidades, and by Consejer\'{\i}a de Econom\'{\i}a, Innovaci\'on, Ciencia y Empleo of Junta de Andaluc\'{\i}a as research group FQM-322, as well as FEDER funds.

%

\vspace{5mm}
\facilities{WEBT, SMARTS, Bok, SO:Kuiper, MMT, 	CASLEO:JST, CAO:2.2m} 


\software{Astropy \citep{2013A&A...558A..33A},
          DAOPHOT \citep{Stetson1987}, IRAF \citep{1986SPIE..627..733T}}





\bibliography{AO0235p164}{}
\bibliographystyle{aasjournal}



\end{document}